\documentclass[journal]{IEEEtran}

\usepackage[utf8]{inputenc}
\usepackage{cite}
\usepackage{overpic}
\usepackage{amsmath,amssymb,amsfonts}
\usepackage{bm}
\usepackage{graphicx}
\usepackage{textcomp}
\usepackage{xcolor}
\usepackage{float}
\usepackage{amsthm}
\usepackage{graphicx}
\usepackage{epstopdf}
\usepackage{amsmath,bm,bbm}
\usepackage{amsfonts}
\usepackage{amssymb}
\usepackage{color}
\usepackage{multirow}
\usepackage{multicol}
\usepackage{soul,xcolor}
\usepackage{setspace}
\usepackage{algorithm}
\usepackage{algpseudocode}
\usepackage[normalem]{ulem}
\newtheorem{lemma}{Lemma}

\definecolor{mypurple}{HTML}{9933FF}
\definecolor{mygreen}{HTML}{009900}

\DeclareMathOperator{\tr}{tr}

\definecolor{orange}{RGB}{0,112,192}
\theoremstyle{plain}

\newcommand{\mathacr}[1]{\mathsf{#1}}

\newcommand{\vect}[1]{\mathbf{#1}}

\def\tr{\mathrm{tr}}

\def\Htran{\mbox{\tiny $\mathrm{H}$}}
\def\Ttran{\mbox{\tiny $\mathrm{T}$}}
\def\CN{\mathcal{N}_{\mathbb{C}}} 
\begin{document}
\makeatletter
\newcommand*{\rom}[1]{\expandafter\@slowromancap\romannumeral #1@}
\makeatother

\title{Cell-Free Massive MIMO in O-RAN: Energy-Aware Joint Orchestration of Cloud, Fronthaul, and Radio Resources 
\thanks{\"O. T. Demir is with the  Department of Electrical and Electronics Engineering, TOBB University of Economics and Technology, Ankara, Turkey (ozlemtugfedemir@etu.edu.tr). M. Masoudi is with the Global AI Accelerator in Ericsson, Stockholm, Sweden (meysam.masoudi@ericsson.com). E.~Bj\"ornson and C. Cavdar are with the Department of Computer Science, KTH Royal Institute of Technology, Kista, Sweden (\{emilbjo, cavdar\}@kth.se).  This work was partially funded by CELTIC-NEXT project AI4Green with the support of Vinnova, Swedish Innovation Agency. The work of Emil Bj\"ornson was supported by the Swedish Foundation for Strategic Research under Grant FFL18-0277.}}
\author{\IEEEauthorblockN{\"Ozlem Tu\u{g}fe Demir, \emph{Member, IEEE}, Meysam Masoudi, \emph{Member, IEEE}, Emil Bj\"ornson, \emph{Fellow, IEEE}, and Cicek Cavdar, \emph{Member, IEEE}}}

\maketitle
\begin{abstract}

For the energy-efficient deployment of cell-free massive MIMO functionality in a practical wireless network, the end-to-end (from radio site to the cloud) energy-aware operation is essential. In line with the cloudification and virtualization in the open radio access networks (O-RAN), it is indisputable to envision prospective cell-free infrastructure on top of the O-RAN architecture. In this paper, we explore the performance and power consumption of cell-free massive MIMO technology in comparison with traditional small-cell systems, in the virtualized O-RAN architecture. We compare two different functional split options and different resource orchestration mechanisms. In the end-to-end orchestration scheme, we aim to minimize the end-to-end power consumption by jointly allocating the radio, optical fronthaul, and virtualized cloud processing resources. We compare end-to-end orchestration with two other schemes: i) ``radio-only'' where radio resources are optimized independently from the cloud and ii) ``local cloud coordination'' where orchestration is only allowed among a local cluster of radio units. We develop several algorithms to solve the end-to-end power minimization and sum spectral efficiency maximization problems. The numerical results demonstrate that end-to-end resource allocation with fully virtualized fronthaul and cloud resources provides a substantial additional power saving than the other resource orchestration schemes.

\end{abstract}

\section{Introduction}

The number of mobile user equipments (UEs) and their capacity requests are anticipated to  increase continuously during the next decade \cite{jiang2021road}. It is expected that by 2030 ubiquitous and limitless connectivity will be needed everywhere \cite{han2018network}. New
networking and new air interface solutions are the two key enablers to support next-generation wireless systems \cite{jiang2021road}. The networking technologies include the softwarization, virtualization, and open radio access networks
(O-RAN), whereas massive MIMO (multiple-input multiple-output) and cell-free massive MIMO are among the main air interface technologies. This paper studies cell-free massive MIMO in the O-RAN architecture, which brings together two inherently compatible networking and air interface entities of beyond 5G networks.

Cell-free massive MIMO  has been proposed as  a physical-layer  technology that combines ultra-dense networks with joint transmission/reception (JT), and the low-complexity linear processing schemes from massive MIMO  \cite{Ngo2017b,interdonato2019ubiquitous,cell-free-book}. By taking advantage of both joint processing and macro diversity, cell-free massive MIMO reduces the large data rate variations across the coverage area, which solves one of the main drawbacks of the current cellular networks. For joint processing of UEs’ signals, it is required to centralize parts of the baseband processing, which constructs the inherent connection between cell-free massive MIMO and the centralized RAN (C-RAN) architecture \cite{wang2020implementation}. The separation of software from hardware  not only allows for joint processing in cell-free operation but also creates new energy-saving opportunities with green and agile virtualization \cite{imoize2022review}. To meet the coordination and signaling requirements of the envisioned cell-free massive MIMO network, O-RAN is envisaged as a promising architecture by providing substantial adaptability \cite{Vardakas2022}.

 In this paper, in a cell-free massive MIMO network with O-RAN architecture, we jointly allocate radio, optical transport network, and cloud processing resources given the number of UEs and their performance requirements in an area. We derive the required processing resources for each operation in the cloud and determine the required optical transport resources based on different functional split options in a cell-free massive MIMO system. The joint orchestration of end-to-end resources enabled by O-RAN architecture is critical to fully benefit from energy-saving mechanisms in a cell-free network. Indeed, O-RAN enables the joint resource allocation by the real-time and near-real-time softwarized controllers \cite{malandrino2022performance}. Thanks to this joint orchestration, the number of active radio units (RUs) can scale down together with the amount of optical and processing resources in the transport network and the cloud, respectively, following the UE demand, to minimize the total end-to-end power consumption of the network.

\subsection{Evolution of RAN architecture towards O-RAN}

The distributed RAN (D-RAN) is the most widely deployed legacy network architecture, in which all baseband processing functions for each base station (BS) are co-located with the RU at the cell site \cite{wang2017virtualized}. This architecture is not scalable, cost-efficient, and, most importantly, is not capable of supporting heterogeneous services efficiently in terms of energy consumption and throughput \cite{alliance2016ngmn}. The C-RAN architecture was developed to improve the network's energy efficiency and resource utilization. In conventional C-RAN, the baseband processing functions are detached from the RUs at the cell site and moved into a centralized resource pool at a central cloud (CC). Although this architecture is more energy-efficient than D-RAN due to centralized cooling, each RU has its dedicated processing unit making the processing resources under-utilized when the traffic is unbalanced between cells during peak hours.

Virtualized C-RAN emerged as an architecture that decouples the hardware and software by virtualizing network functions \cite{wang2017virtualized,wang2016joint,wang2016energy,Masoudi2020,alabbasi2018optimal,sigwele2017energy,masoudi2019green}. In this architecture, the deployed processing units are no longer dedicated to one specific RU, but network functionalities are implemented in software and run on general-purpose processors (GPPs) \cite{bonati2020open}. In this way, processing resources can be shared between various loaded cells, further improving resource utilization and reducing network energy consumption. The advantages of implementing virtualized C-RAN are i) simplified network management; ii) enabled resource pooling; and iii) improved coordination of radio resources required for cell-free operations. Although the virtualized C-RAN architecture advancement is promising, full centralization of physical-layer processing significantly upscales the fronthaul signal capacity, especially when technologies like massive MIMO are employed. Therefore, a more convenient and potentially flexible architecture needed to be further investigated to keep the scalability benefits of C-RAN while resolving the bandwidth congestion and allowing for effective coordinated multipoint (CoMP) coordination for cell-edge UEs. 

 Recently, a consensus has been reached between network vendors and operators to support an O-RAN architecture and standards \cite{garcia2021ran}. The O-RAN Alliance \cite{alliance2019ran} is an industry-wide standardization for RAN interfaces that complements the 3GPP standards and covers RAN disaggregation, RAN automation, and RAN virtualization.  O-RAN architecture, proposed by the O-RAN Alliance, is a virtualized C-RAN with an open, interoperable interface and virtualization, allowing multiple vendor products to work together in one network. O-RAN and following standardization efforts enable building the virtualized C-RAN on open hardware and cloud, and allowing full exploitation of virtualization and sharing in virtualized C-RAN. The three key elements of O-RAN are i) cloudification; ii) intelligence and automation; and iii) open internal RAN interface. The primary mission of the O-RAN is to reshape the RAN industry towards open, virtualized, and fully interoperable mobile networks \cite{polese2022understanding}.

\subsection{Cell-free massive MIMO in the O-RAN architecture}

In the previous work, mostly the physical-layer aspects of cell-free massive MIMO have been studied and only the radio site power consumption has been considered \cite{van2020joint}. In \cite{agheli2020designing}, the power consumption of the fronthaul transport is also taken into account, but the authors assume all RUs serve all UEs, which is not power efficient. 
In \cite{wang2017virtualized}, a cell-free network architecture is presented by optimizing the user-centric formation of soft BSs defined as joint allocation of spectrum, optical wavelength, and cloud processing resources together with a set of RUs considering JT. End-to-end power consumption\cite{wang2017virtualized} and network throughput\cite{wang2016joint} are optimized considering radio, optical fronthaul, and cloud processing resources. However, massive MIMO is not considered in these cell-free networks. In \cite{pan2017joint}, RU selection for JT under fronthaul constraints is studied, but the cloud processing power consumption is simplified as a fixed parameter. In \cite{ha2016computation}, the processing requirements in the cloud are taken into consideration, but only radio site power consumption is minimized.

Recently, the authors of \cite{ranjbar2022cell} have refined the cell-free massive MIMO terminology according to the O-RAN architecture and discussed several implementation options of cell-free functionality in the current or future O-RAN generations.  The works \cite{vardakas2021towards,Vardakas2022} also studied the performance of the cell-free massive MIMO on top of the O-RAN architecture. In \cite{murakami2022analysis}, the placement of the central processing unit and allocation  of radio bandwidth have been studied in terms of throughput. 
To the best of the authors' knowledge, all the related work on cell-free massive MIMO in O-RAN either considered architectural high-level views or focused on spectral efficiency (SE) or throughput performance.

\subsection{Contributions}

In the conference version \cite{Demir2022ICC} of this paper, the end-to-end power consumption modeling and minimization were considered only with fully-centralized functional splitting option in virtualized C-RAN architecture. In this paper, we follow a holistic approach by studying different resource orchestration schemes for the cell-free massive MIMO and small-cell systems in the O-RAN architecture. The considered end-to-end network power consumption for cell-free massive MIMO involves the impact of radio, optical fronthaul, and cloud processing resources. Extending the conference version, we additionally consider intra-physical-layer functional splitting option by modifying the power consumption accordingly. We derive the cloud processing requirements of a cell-free massive MIMO OFDM system, given the required system performance. Based on our developed end-to-end power consumption model, we cast two optimization problems to jointly allocate the radio, fronthaul, and cloud resources. The first problem, which was the only problem considered in \cite{Demir2022ICC}, minimizes the end-to-end power consumption by joint allocation of transmit powers, optical fronthaul resources, cloud processing resources, and the set of O-RUs (in line with the O-RAN terminology) serving the UEs to meet their quality of service (QoS) requirements. We cast the problem in a mixed binary second-order cone programming form, which can be optimally solved. Different from the conference paper, an approximated version of the original power minimization problem is obtained via $l_0$ norm, and a concave-convex programming (CCP)-based algorithm is  proposed to solve this problem in a more manageable form to gain further insights into larger cell-free setups. Moreover, a joint sum SE maximization and total network power minimization problem is proposed. After novel transformations, a proper approximated form of this problem is solved via the same CCP-based algorithm. Using the found solutions, we compare the performance of the fully virtualized end-to-end, local cloud coordination-based, and radio-only resource allocation, where only the end-to-end scheme was considered in \cite{Demir2022ICC}. Through numerical simulations, we show how much power saving is achieved by the virtualized end-to-end resource allocation compared to the case of fixed fronthaul resources and partial resource sharing in the cloud (local coordination), and the cloud-unaware radio-only scheme. Moreover, the SE improvement provided by the cell-free massive MIMO over conventional small-cell networks, where each UE is served by only one O-RU, is quantified for different scenarios. The effect of different functional splits is discussed.

\subsection{Paper Outline}
The remainder of this paper is organized as follows. Section~\ref{sec:architecture} overviews the O-RAN architecture for cell-free massive MIMO functionality. Section~\ref{sec:system} introduces the channel model, channel estimation, and downlink operation in a cell-free massive MIMO system. In Section~\ref{sec:power}, the end-to-end power consumption modeling together with the analysis of processing complexity is elaborated. The details of the end-to-end power minimization problem and the respective algorithms are provided in Section~\ref{sec:optimization}. Section~\ref{sec:sumrate} extends the developed optimization methodology to the joint sum SE maximization and power minimization problem. The performance of the proposed end-to-end energy-aware algorithms is quantified and compared with the partial energy-saving mechanisms in Section~\ref{sec:numerical}. Finally, Section~\ref{sec:conclusions} concludes the paper.    

{\bf Reproducible research:} All the simulation results can be reproduced using the Matlab code available at:    https://github.com/ozlemtugfedemir/O-RAN-cell-free

\section{Architecture Overview of O-RAN for Cell-free Massive MIMO}\label{sec:architecture}

\begin{figure*}[t]
	\centering
	\begin{overpic}[width=14cm,tics=10]{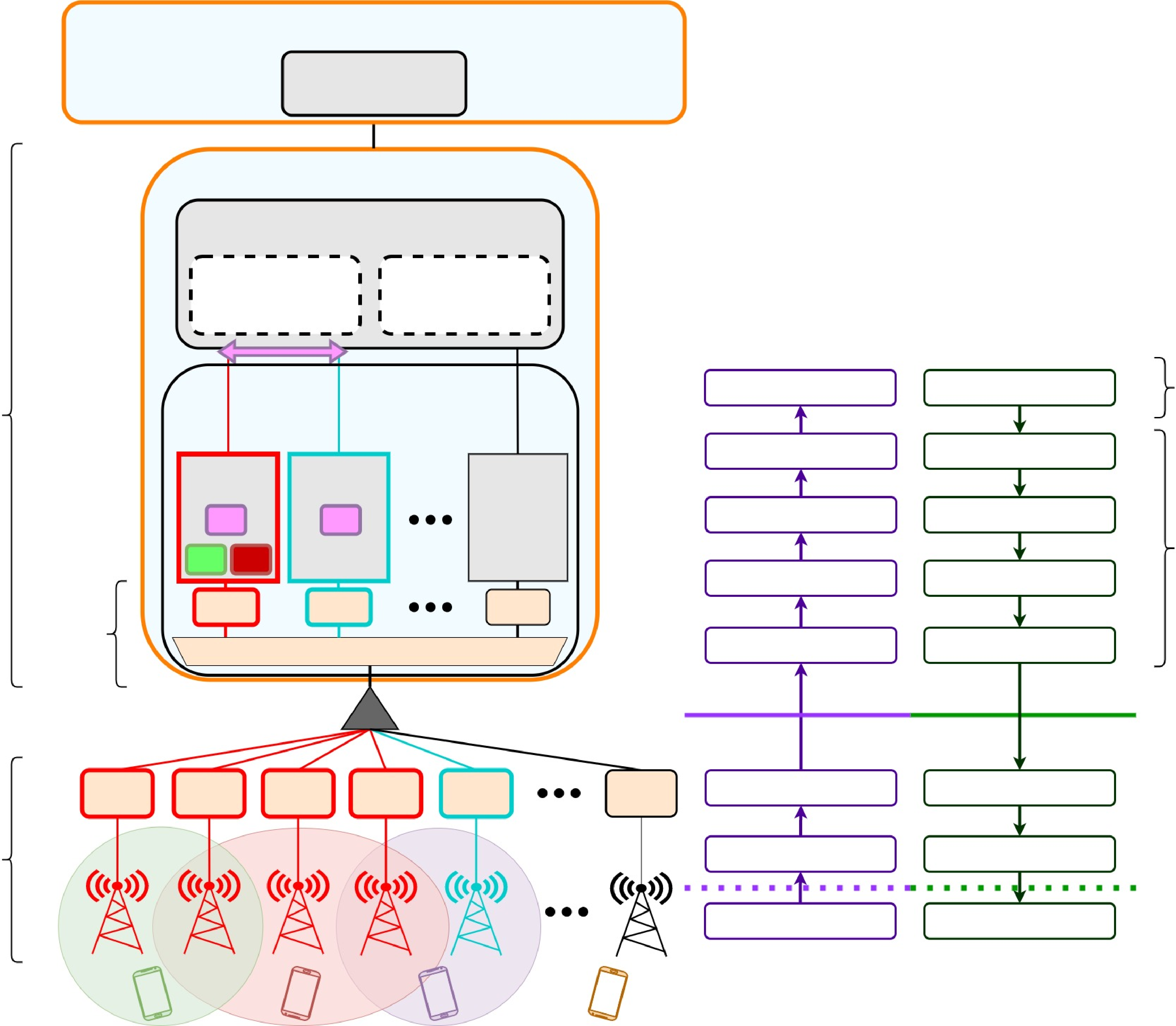}
		\put(10.6,-1.8){\small UE $1$}				\put(22.6,-1.8){\small UE $2$}
		\put(35,-1.8){\small UE $3$}
		\put(48.6,-1.8){\small UE $K$}
		\put(2.1,14.1){\small O-RU\,$1$}
		\put(10.4,14.1){\small O-RU\,$2$}
		\put(46.5,14.1){\small O-RU\,$L$}
		\put(15.5,46.5){\small GPP\,$1$}	
		\put(24.9,46.5){\small GPP\,$2$}
		\put(40,46.5){\small GPP\,$W$}
		\put(7.5,18.9){\small ONU}
		\put(15.3,18.9){\small ONU}
		\put(22.9,18.9){\small ONU}
		\put(30.3,18.9){\small ONU}
		\put(38,18.9){\small ONU}
		\put(52.1,18.9){\small ONU}
		\put(17.8,34.8){\small LC}
		\put(27.5,34.8){\small LC}
		\put(42.65,34.8){\small LC}
		\put(4.4,32.5){\small OLT}
		\put(-8.6,51.2){\small O-Cloud}
		\put(-10,13.5){\small Radio site}
		\put(-10.2,25.4){\small eCPRI-based fronthaul/midhaul}
		\put(34,26.5){\small Optical splitter}
		
		\put(25.5,31.05){\small WDM MUX}
		\put(9,84.5){ \small Service management and orchestration (SMO)}
		\put(25,79.5){ \small Non-RT RIC}
		\put(26,71.6){ O-Cloud}
		\put(30,52){\small O-CU+O-DU}
		\put(24.7,67){ \small Near-RT RIC}
		\put(20.5,62.5){ \small QoS}
		\put(16.7,60){ \small management}
		\put(36,63.4){ \small Radio }
		\put(33.8,61.5){ \small connection }
		\put(33,59.5){ \small management}
		\put(97,10.8){ \color{red}{Option 8}}
		\put(97,25.45){ \color{red}Option 7.2}
		\put(64,4.7){ \small Uplink}
		\put(81.2,4.7){ \small Downlink}
		\put(66.6,8.1){\small RF}
		\put(85.2,8.1){\small RF}
		\put(61.1,13.8){\small CP rem.+DFT}
		\put(80.3,13.8){\small IDFT+CP ins.}
		\put(61.2,19.45){\small RE demapping}
		\put(80.95,19.45){\small RE mapping}
		\put(80.5,31.55){\small Tx precoding}
		\put(61.6,31.55){\small Rx combining}
		\put(61.55,37.25){\small Demodulation}
		\put(81.35,37.25){\small Modulation}
		\put(83.2,42.7){\small Coding}
		\put(63.5,42.7){\small Decoding}
		\put(62.7,48.05){\small MAC+RLC}
		\put(81.2,48.05){\small MAC+RLC}				\put(62.35,53.45){\small Higher layer}
		\put(80.9,53.45){\small Higher layer}
		\put(100.3,53.45){\small O-CU}
		\put(100.3,39.8){\small O-DU}
	\end{overpic}
	\caption{O-RAN architecture for cell-free massive MIMO with functional splitting options 8 and 7.2 in uplink and downlink.}
	\label{fig:architecture}
\end{figure*}
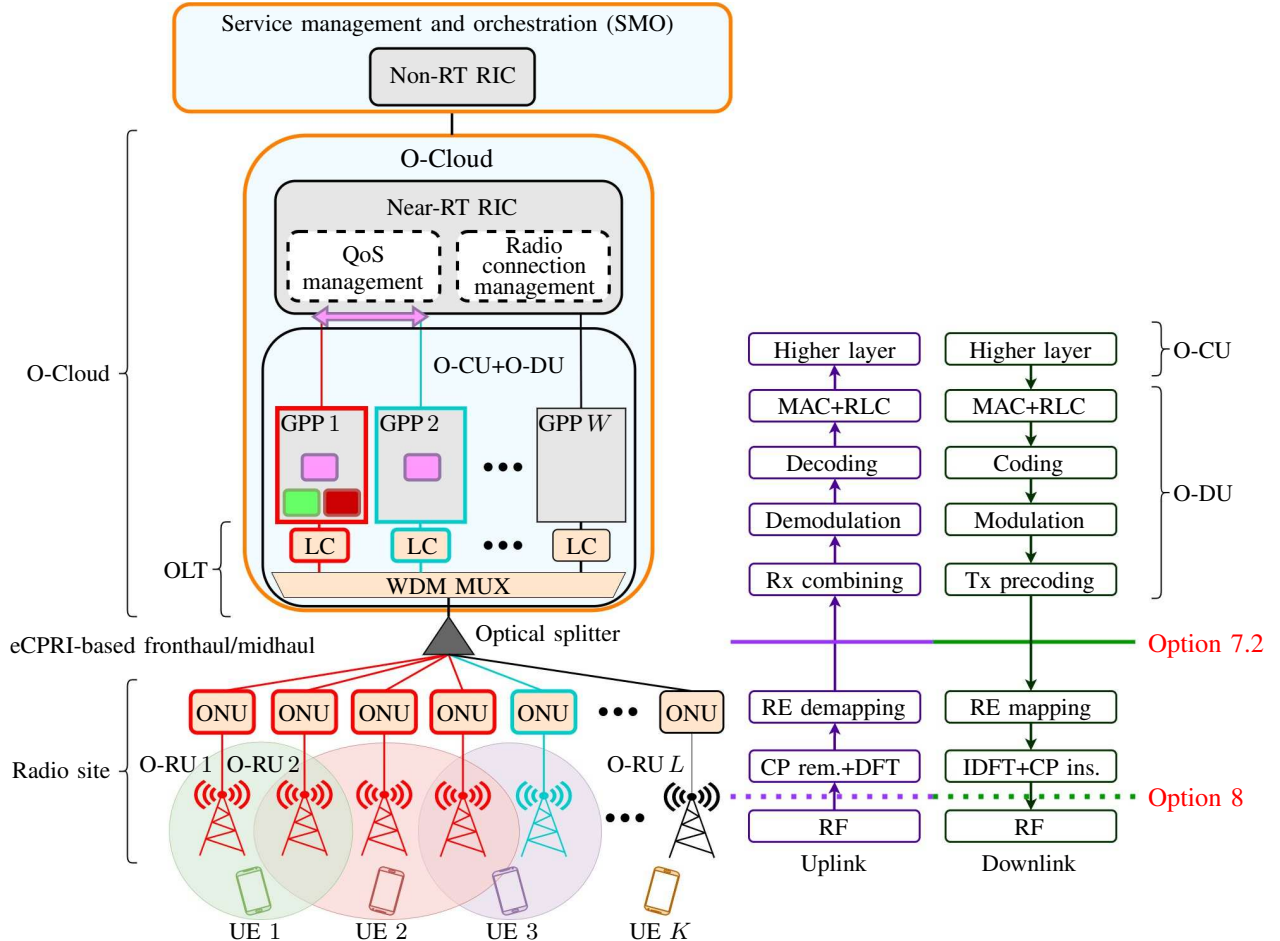

We consider a cell-free massive MIMO system that is built on the top of the O-RAN architecture in line with the next-generation virtualized C-RAN ecosystem as shown in Fig.~\ref{fig:architecture} \cite{garcia2021ran}. There are $L$ O-RUs and $K$ UEs that are arbitrarily distributed in the coverage area. All UEs have a single antenna while each O-RU is equipped with $N$ antennas. All the O-RUs are connected to the O-Cloud with virtualization and processing resource sharing capabilities  \cite{wang2017virtualized}, via fronthaul connections. O-Cloud consists of two main units, which are O-CU (centralized unit) and O-DU (distributed unit). According to the O-RAN specification, the O-DU is responsible for the lower network layer operations (RCL, MAC, and PHY) whereas the O-CU implements the higher layer operations as illustrated in Fig.~\ref{fig:architecture}.  O-RAN also has logical nodes known as near real-time RAN intelligent controller (near-RT RIC) and non-RT RIC. The near-RT-RIC is responsible for  near real-time intelligent optimization of RAN resources. On the other hand, non-RT RIC is located in the service management and orchestration (SMO) unit, which is responsible for non-real-time intelligent orchestration. These two logical nodes enable fully virtualized end-to-end resource optimization, which we consider in this paper.  O-RAN has multiple deployment options, in some of which O-DU and O-CU are co-located, and in some of which, they are separated logically and geographically.

 In this study, we consider Scenario A as the O-RAN deployment setup,  where O-CU and O-DU are bundled and co-located with the logical node near-RT RIC communicating through E2 interface defined by O-RAN \cite{oran-white-paper}. One key advantage of this scenario is the minimized delay while the deployment cost might be larger compared to the other scenarios \cite{Dryjanski_2023}. Hierarchical deployment scenarios in O-RAN are kept as future work since they will not affect the key findings of our study. A set of O-RUs serves each UE coherently to improve SE. Let $x_{k,l}\in\{0,1\}$ be the binary variable denoting whether UE $k$ is served by O-RU $l$ or not. It is one if UE $k$ is served by O-RU $l$, and zero otherwise. 
For example, in Fig.~\ref{fig:architecture}, the colored circular regions for each UE indicate the O-RUs that are serving them. In this paper, we will optimize these subsets to satisfy several optimization metrics that take the SE of each UE and the end-to-end network power consumption into account.

In the O-Cloud, there are $W$ stacks of general-purpose processors (GPPs) in line with the cloudification framework of O-RAN \cite{garcia2021ran}. These pooled GPPs are used for baseband processing due to their processing capabilities and programmability, which allows virtualization.  The workload of each GPP is routed via a dispatcher that is controlled by a global cloud controller \cite{sigwele2017energy}, which is near-RT-RIC in the considered O-RAN architecture. To perform cell-free joint processing, the respective data and control signals for a particular UE should first exist in the same GPP even though the locally precoded signals can be computed and sent to the O-RUs from different GPPs (thanks to distributed operation and computational sharing and virtualization among multiple GPPs).
In Fig.~\ref{fig:architecture}, the same colors are used to show which O-RUs are connected to which GPP. UE 1 is served by O-RU 1 and O-RU 2, which are all connected to GPP 1. In this case, UE 1 can be served using potentially only GPP 1 without any need for data exchange among GPPs. However, since the O-RUs that serve UE 3 are connected to different GPPs, the connection should be activated between GPP 1 and GPP 2 for the sharing of UE 3 data and control signal. These are all orchestrated by the near-RT-RIC.

We consider the recently released evolved CPRI (eCPRI) specification for the fronthaul/midhaul transmission. A time- and wavelength-division multiplexed passive optical network (TWDM-PON) is utilized as the fronthaul transport network to carry eCPRI packets to meet the high capacity requirements of the fronthaul transmission in a cell-free network \cite{wang2017virtualized,wang2016energy}. As shown in Fig.~\ref{fig:architecture}, each O-RU is connected to one optical network unit (ONU) that is assigned to one of the multiple wavelengths in the fiber communication. Each wavelength can be shared by more than one O-RU using time-division multiplexing. In the O-Cloud, there exists an optical line terminal (OLT) with a WDM multiplexer (WDM MUX) and multiple line cards (LCs), each of which is connected to one GPP. Each LC serves only one wavelength and, thus, each O-RU's signals are received at or transmitted from the GPP that uses the same wavelength. For example, GPP 1 is responsible for the fronthaul transport of the first four O-RUs' data, whereas O-RU 5 is connected to GPP 2.

In the considered O-RAN architecture, two different functional split options from 3GPP specification  \cite{3GPP_functional_split} are considered. The first one, which is called \emph{Option 8} and shown in Fig.~\ref{fig:architecture}, is the physical layer (PHY)-radio frequency (RF) functional split to fully benefit from the efficient processing resources and housing facilities in the O-Cloud.\footnote{The functional split option 7.2 is the one that is mainly supported by O-RAN. However, functional split option 8 is also important to consider not only due to its energy-saving potential but also the high experimentation interest by the leading researchers in the field according to the survey results in \cite[Fig.~1]{abdalla2022toward}.}  According to this split, the O-RUs only perform RF processing and transmit (receive) the pure sampled and quantized baseband signals to (from) the O-Cloud via fronthaul links. All the remaining processing is done in the O-Cloud. Hence, Option 8 has the advantage of the lowest RU complexity \cite{rodriguez2020cloud}. In particular, with  functional split option 8 in the uplink training, all the operations below the dashed purple line in Fig.~\ref{fig:architecture} are implemented at the O-RUs whereas the remaining physical-layer, MAC and RLC operations are carried out in the O-DU. Then, higher-layer functions are implemented in O-CU.  Similarly, the dashed green line determines the same functional split for the downlink operation. With PHY-RF functional split option 8, the required fronthaul data rate for each AP is given as \cite{perez2018fronthaul}
\begin{equation}
    R_{\rm fronthaul}^{(8)} = 2f_sN_{\rm bits}N,
\end{equation}
where $f_s$ and $N_{\rm bits}$ are the sampling frequency and the number of bits to quantize the signal samples, respectively. Due to the limited capacity of each wavelength in TWDM-PON, which is denoted by $R_{\rm max}$, we can assign at most $W_{\rm max}\triangleq\lfloor R_{\rm max}/  R_{\rm fronthaul}^{(8)}\rfloor$ O-RUs to each wavelength and, hence, to each GPP $w$, for $w=1,\ldots,W$.  

The other functional split is \emph{Option 7.2} and is more specifically considered in O-RAN \cite{rodriguez2020cloud,o-ran_fronthaul}. This option splits the network functions so that some of the PHY operations close to RF processing (low-PHY functions) are implemented in the O-RUs whereas the remaining PHY (high-PHY) and higher layer operations are moved to the O-Cloud as shown by the solid purple and green lines in Fig.~\ref{fig:architecture}.\footnote{ In the literature, some authors use the term \emph{midhaul} to refer to the link between O-RU and O-DU when the functional split option 7.2 is adopted. In this paper, we use \emph{fronthaul} to refer to the corresponding link for all the split options without loss of generality.} This functional splitting option generally lowers the fronthaul data rate compared to Option 8, by still allowing cell-free JT processing and low O-RU complexity. The uplink receive combining and transmit precoding are considered in the O-Cloud for coherent JT. The corresponding fronthaul data rate requirement is 
given as \cite{perez2018fronthaul}
\begin{equation}
    R_{\rm fronthaul}^{(7.2)} = \frac{2N_{\rm bits}N_{\rm used}N}{T_s},
\end{equation}
where $N_{\rm used}$ is the number of used subcarriers and $T_s$ is the OFDM symbol duration.  We expect Option 7.2 is more advantageous since it  usually holds that  $R_{\rm fronthaul}^{(7.2)}< R_{\rm fronthaul}^{(8)}$ so that we can assign more O-RUs to each GPP, i.e., $W_{\rm max}\triangleq\lfloor R_{\rm max}/  R_{\rm fronthaul}^{(7.2)}\rfloor$ O-RUs to each wavelength. To quantify this, let us consider a setup with $R_{\rm max}=10$\,GBps, $N_{\rm bits}=12$, and conventional LTE parameters $f_s=30.72$\,MHz, $T_s=71.4$\,$\mu$s $N_{\rm used}=1200$. Further assuming $N=4$, each GPP, and, thus each wavelength, can support $W_{\rm max}=3$ O-RUs with Option 8,  while $W_{\rm max}=6$ O-RUs can be connected to each GPP with Option 7.2. When the number of O-RU antennas is greater than 13, the maximum capacity of each TWDM-PON wavelength is not big enough to provide a fronthaul connection between the O-RUs and the O-Cloud with Option 8. However, with Option 7.2, still, up to 24-antenna O-RUs can be supported by each wavelength due to the reduced fronthaul rate requirement.

Apart from fronthaul transport limitations, the relation of the processing requirements in the O-Cloud and how the GPPs can handle those based on their limited capabilities determine the number of active GPPs. Shutting down the unused active GPPs and the corresponding LCs, we can save power by minimizing the active idle power of the network \cite{sigwele2017energy}. However, there is a trade-off between minimizing the number of active GPPs and the SE performance of UEs. In this paper, we will derive the related processing requirements of each operation in the O-Cloud for a cell-free massive MIMO OFDM system. Before we introduce our optimization problems, we will present the physical-layer foundations of cell-free massive MIMO system in the next section.

\section{Cell-Free Massive MIMO System}\label{sec:system}

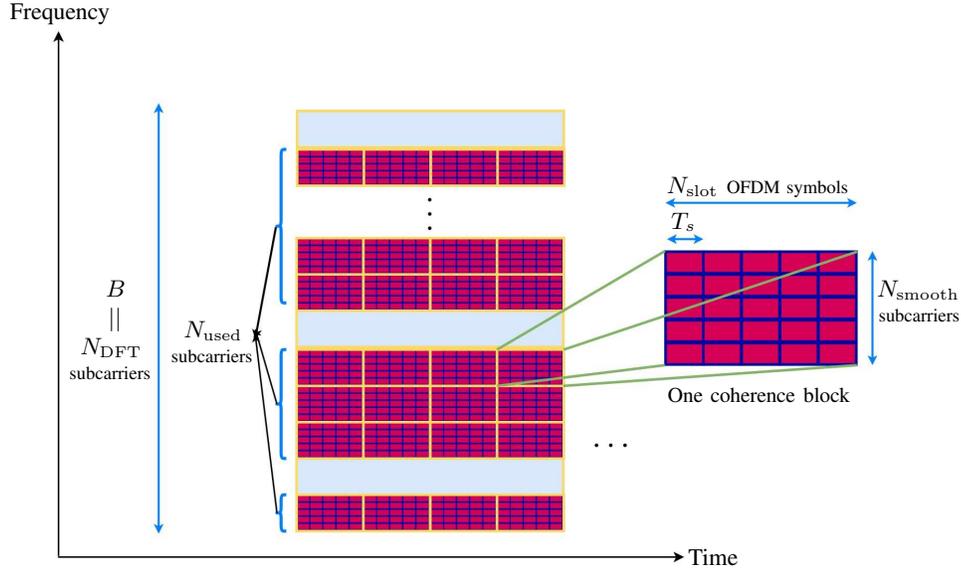
\begin{figure*}[t]
\centering
    \begin{overpic}[width=11cm,tics=10]{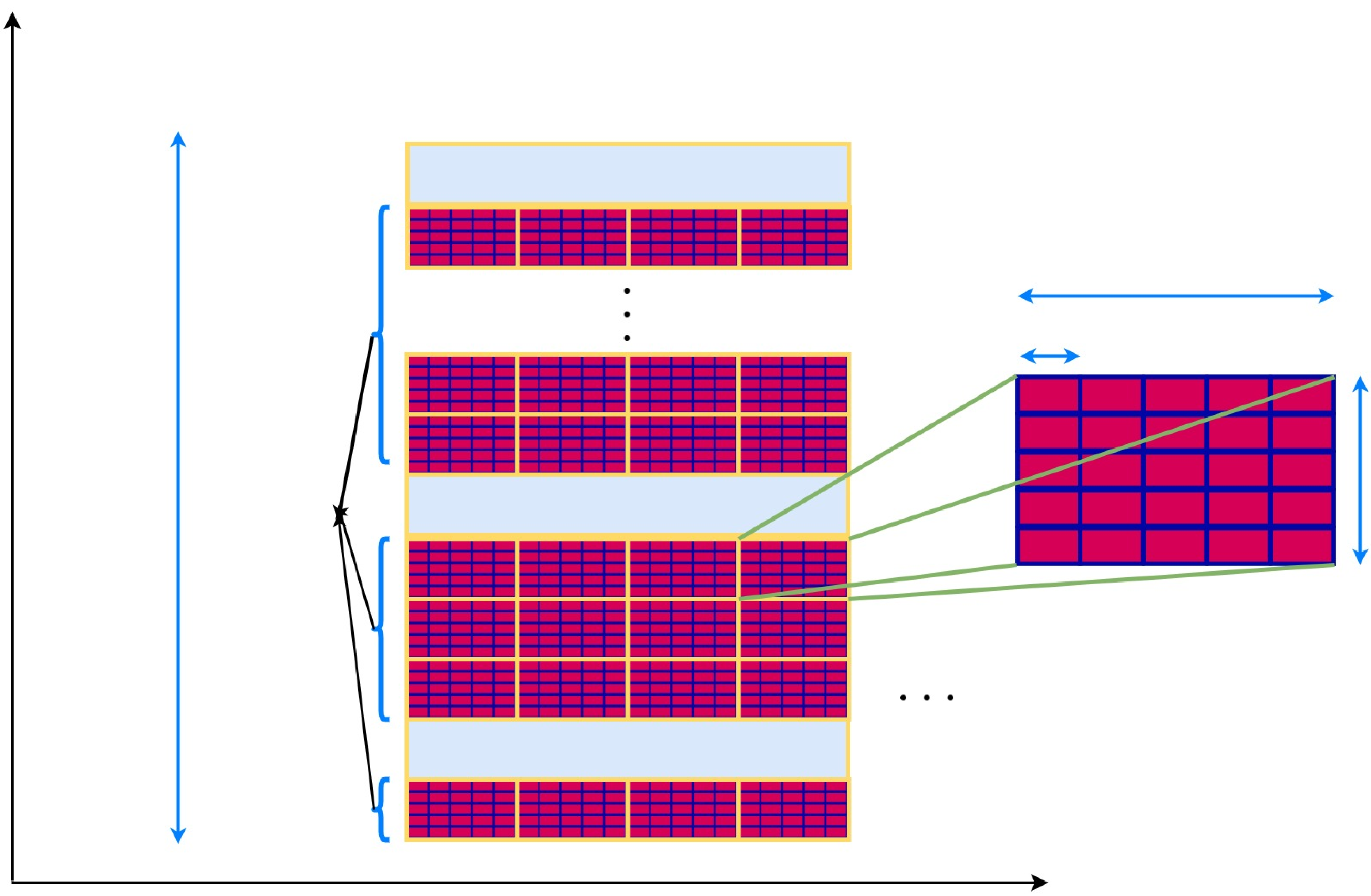}
      \put(6.6,32.4){\small $B$}
       \put(6.9,29){\small $|  |$}
            \put(3.5,25.5){\small $N_{\rm DFT}$}
            \put(2.2,22.5){\scriptsize subcarriers}
    \put(16,27.5){\small $N_{\rm used}$}	
            \put(14.5,24.5){\scriptsize subcarriers}
		          \put(99.8,32.5){\small $N_{\rm smooth}$}	
        \put(100,29.5){\scriptsize subcarriers}	
                    \put(74.4,45){\small $N_{\rm slot}$ \scriptsize OFDM symbols}
                   \put(74.9,40.7){\small $T_s$}
                    \put(77,0){\small Time}
                    \put(-5,66){\small Frequency}
                    \put(74.5,19.7){\footnotesize One coherence block}
		\end{overpic}
	\caption{In each coherence block of $\tau_c= N_{\rm smooth}N_{\rm slot}$ complex samples, the channel is modeled as time-invariant and frequency-flat according to the  block-fading model. Out of total number of $N_{\rm DFT}$ subcarriers, $N_{\rm used}\leq N_{\rm DFT}$ subcarriers are utilized.}
	\label{fig:block-fading}
\end{figure*}

A cell-free massive MIMO system with time-division duplex and OFDM is considered. The carrier and sampling frequencies are $f_c$ and $f_s$, respectively. We assume a block-fading channel model as illustrated in Fig.~\ref{fig:block-fading}. The total number of subcarriers is $N_{\rm DFT}$ across the total bandwidth of $B$\,Hz. $N_{\rm DFT}$ is also the dimension of the discrete Fourier transform (DFT), while the number of used subcarriers is $N_{\rm used}\leq N_{\rm DFT}$ shown as red rectangles. The light blue rectangles represent the null subcarriers. Each OFDM symbol has a duration of  $T_{ s}$ seconds. According to the block fading channel modeling, the channels are constant time-invariant and frequency-flat in each coherence block that consists of $N_{\rm smooth}$ consecutive OFDM subcarriers and  $N_{\rm slot}$ OFDM symbols\cite{Marzetta2016a}, \cite[Remark 2.1]{cell-free-book}. The channel is constant across $\tau_c=N_{\rm smooth}N_{\rm slot}$ channel uses, which is the number of useful samples in each coherence block, and takes independent realizations between different blocks. Although in practice each OFDM sample (channel use) has a unique channel response, the block-fading assumption can be accurately applied with a proper selection of $N_{\rm slot}$ and $N_{\rm smooth}$ since the channels do not change abruptly across consecutive time and frequency samples or there is a known mapping between them \cite[Remark 2.1]{cell-free-book}. Hence, the channels can be assumed to be approximately constant (smooth) across $\tau_c$ time-frequency samples in each coherence block without loss of generality.

 There are mainly two types of cell-free network operation. The first one is the centralized operation, where all the processing regarding channel estimation and payload data detection/precoding are performed in the central processing unit (O-Cloud in the considered architecture) \cite[Sec. 5.1 and 6.1]{cell-free-book}. On the other hand, in the distributed operation \cite[Sec. 5.2 and 6.2]{cell-free-book}, each O-RU first estimates the local channels and these estimates are used for local combining/precoding during data transmission. In this paper, we consider distributed downlink operation where the channel estimates corresponding to each O-RU are used for distributed per-O-RU precoding \cite[Sec. 6.2]{cell-free-book}.  In this way, we can divide the UE-related processing tasks into small independent blocks that can be virtualized in the O-Cloud to minimize the number of active GPPs.\footnote{This is not possible in the centralized cell-free operation where higher-dimensional precoding should be computed jointly for all the serving O-RUs, based on pooling of channel estimates, and then applied to a particular UE in the same GPP.} Note that the fronthaul requirements scale with the number of O-RU antennas with both functional split Options 8 and 7.2. In addition, it is proportional to  the number of used subcarriers if Option 7.2 is adopted. In both options, they do not depend on the number of UEs that an O-RU can serve.

Since we focus on the downlink operation, each coherence
block is divided into two phases: i) uplink training with $\tau_p$
samples and ii) downlink payload data transmission with $\tau_d =\tau_c - \tau_p$ samples. All the UEs are served on the same time-frequency resources using spatial multiplexing. Moreover, channel estimation and precoding are implemented in each coherence block in the same way. Hence, we will focus on an arbitrary time-frequency
resource block as in \cite{cell-free-book}.

\subsection{Channel Model}

We let $\vect{h}_{kl}\in \mathbb{C}^{N}$ denote the frequency-domain channel from UE $k$ to O-RU $l$ in an arbitrary coherence block. The channels are modeled using correlated Rayleigh fading, i.e.,  $\vect{h}_{kl}\sim\mathcal{N}_{\mathbb{C}}({\bf 0}_N,{\bf R}_{kl})$ and they are independent for different UEs and O-RUs. The correlation matrix ${\bf{R}}_{kl}\in \mathbb{C}^{N \times N}$ is determined by the spatial correlation of the channel $\vect{h}_{kl}$ between the antennas of O-RU $l$ and the corresponding average channel gain, which is denoted by $\beta_{kl}=\tr({\bf R}_{kl})/N$. The channel gain is dependent on large-scale effects such as geometric attenuation and shadowing. The spatial correlation matrices are fixed throughout the communication and they are known in accordance with the related literature \cite{cell-free-book, MassiveMIMO20}.

\subsection{Uplink Training: Channel Estimation}

        To perform coherent transmit  processing, the channels need to be estimated in each coherence block. In a large cell-free network with many UEs, there will not be enough pilot resources to assign orthogonal sequences to all UEs. Hence, we consider a set of $\tau_p$ mutually orthogonal pilot sequences $\boldsymbol{\phi}_{1},\ldots,\boldsymbol{\phi}_{\tau_p}\in\mathbb{C}^{\tau_p}$ that are assigned to the UEs and reused by multiple UEs. The sequences satisfy
     \begin{equation}
\boldsymbol{\phi}_{t_1} ^{\Htran} \boldsymbol{\phi}_{t_2} = \begin{cases} \tau_p, & t_1 = t_2,\\
0, & t_1 \neq t_2. \end{cases}
\end{equation}    
The pilot sequences are assigned to the UEs in a deterministic way and $t_k$ denotes the index of the pilot assigned to UE $k$ as $t_k \in \{ 1, \ldots, \tau_p\}$. The set of UEs that share the same pilot with UE $k$ is defined as
\begin{equation}
\mathcal{P}_k = \left\{ i : \  t_i = t_k, \ i=1,\ldots,K  \right\} \subset \{ 1, \ldots, K\}.
\end{equation}
As shown in Fig.~\ref{fig:architecture}, either after initial RF processing (with Option 8) or after resource demapping (with Option 7.2), the received signals are transmitted to the O-Cloud and the channel estimation is performed there for each coherence block. The frequency-domain received signal at O-RU $l$ and a particular coherence block during the entire pilot transmission is given by
\begin{equation} \label{eq:received-pilot-matrix}
\vect{Y}_{l}^{{\rm pilot}} = \sum_{i=1}^{K} \sqrt{\eta_i } \vect{h}_{il} \boldsymbol{\phi}_{t_i}^{\Ttran}+ \vect{N}_{l} 
\end{equation}
where $\eta_i\geq 0$ is the pilot transmit power of UE $i$ and $\vect{N}_{l}  \in \mathbb{C}^{N \times \tau_p}$ is the receiver noise with independent and identically distributed (i.i.d.) elements as $ \CN (0, \sigma^2)$. Using all the received pilots signals, the minimum mean-squared error (MMSE) channel estimate $\widehat{\vect{h}}_{kl}$ of the channel $\vect{h}_{kl}$ is given as follows \cite[Corol.~4.1]{cell-free-book}:
\begin{equation} \label{eq:estimates}
\widehat{\vect{h}}_{kl} = \sqrt{\eta_k} \vect{R}_{kl} \left(\sum_{i \in \mathcal{P}_k} \eta_i\tau_p \vect{R}_{il} + \sigma^2 \vect{I}_{N}\right)^{-1} \vect{Y}_{l}^{{\rm pilot}}  \boldsymbol{\phi}_{t_k}^{*}. 
\end{equation}

 \subsection{Downlink Data Transmission}

Let $\varsigma_i\in \mathbb{C}$ denote the downlink data signal of UE $i$ with $\mathbb{E} \{|\varsigma_i|^2 \} = 1$. Let the normalized (in terms of average power) precoding vector and transmit power corresponding to UE $i$ and O-RU $l$ for $x_{i,l}=1$ be $\vect{w}_{il}\in \mathbb{C}^N$ and $p_{il}\geq 0$, respectively.  In the O-Cloud, the frequency-domain precoded signal to be transmitted from O-RU $l$ is constructed as
\begin{equation} \vect{x}_l=\sum_{i=1}^K\sqrt{p_{il}}x_{i,l}\vect{w}_{il}\varsigma_i\in \mathbb{C}^{N}.
\end{equation}
 In selecting the precoding vectors (local partial MMSE (LP-MMSE) precoding \cite[Sec. 6.2.2]{cell-free-book}), the MMSE channel estimates $\{\widehat{\vect{h}}_{il} \}$ from \eqref{eq:estimates}  are used.
The received frequency-domain downlink signal at UE $k$ is\footnote{Note that an additional conjugation on the channel vectors is not introduced different from \cite{cell-free-book}, so the Hermitian transpose on the channel vectors in   \cite{cell-free-book} are replaced by tranpose.}
\begin{equation}  \label{eq:downlink-received-UEk}
y_k^{\rm{dl}} = \sum_{l=1}^{L}  {\vect{h}}_{kl}^{\Ttran} \vect{x}_l + n_k =\sum_{l=1}^{L} \sum_{i=1}^{K}\sqrt{p_{il}}  x_{i,l} {\vect{h}}_{kl}^{\Ttran} \vect{w}_{il} \varsigma_i + n_k 
\end{equation}
where $n_k \sim \CN(0,\sigma^2)$ is the receiver noise. The downlink achievable SE of UE $k$ at each resource block can be computed using \cite[Corr.~6.3 and Sec. 7.1.2]{cell-free-book} as follows:
	\begin{equation} \label{eq:downlink-rate-expression-level2}
	\mathacr{SE}_{k} = \frac{\tau_d}{\tau_c} \log_2  \left( 1 + \mathacr{SINR}_{k}  \right) \quad \textrm{bit/s/Hz}
	\end{equation}
	with the effective signal-to-interference-plus-noise ratio (SINR) given by
\begin{equation}\label{eq:SINR-downlink-2}
\mathacr{SINR}_{k}=\frac{\left|\vect{b}_k^{\Ttran}{\boldsymbol{\rho}}_k\right|^2}{\sum\limits_{i=1}^K{\boldsymbol{\rho}}_i^{\Ttran}{\vect{C}}_{ki}{\boldsymbol{\rho}}_i+\sigma^2}
\end{equation}
where
\begin{align}
 &{\boldsymbol{\rho}}_k=\left [ \sqrt{p_{k1}}x_{k,1} \, \ldots \, \sqrt{p_{kL}} x_{k,L} \right ]^{\Ttran} \in \mathbb{R}_{\geq 0}^L,  \\
  &{\vect{b}}_k \in \mathbb{R}_{\geq 0}^L, \quad
    \left[ {\vect{b}}_{k} \right]_{l}= \mathbb{E} \left\{ \vect{h}_{kl}^{\Ttran}{\vect{w}}_{kl}\right\}\\
  & {\vect{C}}_{ki}\in \mathbb{C}^{L \times L}, \nonumber \\
  & \left[ {\vect{C}}_{ki} \right]_{lr}=\begin{cases}
  \mathbb{E} \left\{ \vect{h}_{kl}^{\Ttran}{\vect{w}}_{kl}{\vect{w}}_{kr}^{\Htran}\vect{h}_{kr}^*\right\}- \left[ {\vect{b}}_{k} \right]_{l} \left[ {\vect{b}}_{k} \right]_{r}^*, & i=k,  \\
  \mathbb{E} \left\{ \vect{h}_{kl}^{\Ttran}{\vect{w}}_{il}{\vect{w}}_{ir}^{\Htran}\vect{h}_{kr}^*\right\}, & i\neq k. \end{cases} 
\end{align}

In this paper, we use a slightly modified version of the LP-MMSE precoding \cite[Sec. 6.2.2]{cell-free-book}, where $\vect{w}_{kl}=\overline{\vect{w}}_{kl}/\sqrt{\mathbb{E}\{\|\overline{\vect{w}}_{kl}\|^2\}}$ with 
\begin{align} \overline{\vect{w}}_{kl}=\eta_k\left(\sum_{i\in\overline{\mathcal{P}}_l}\eta_i\left(\widehat{\vect{h}}_{il}\widehat{\vect{h}}_{il}^{\Htran}+\vect{Z}_{il}\right)+\sigma^2\vect{I}_N\right)^{-1}\widehat{\vect{h}}_{kl}.
\end{align}
Here, $\vect{Z}_{il}$ is the correlation matrix of the channel estimation error $\vect{h}_{il}-\widehat{\vect{h}}_{il}$ and $\overline{\mathcal{P}}_l$ only includes the UE indices with the strongest channel per each pilot at O-RU $l$.

\section{Power Consumption Modeling}\label{sec:power}

For the considered virtualized O-RAN architecture, we do not stick to specific hardware but instead adopt a generic power model that can be applied to the different technologies. The network power consumption consists of two main components: i) the radio site power consumption that includes the O-RU power consumption $P_{{\rm RU},l}$, for $l=1,\ldots,L$ and the ONU power consumption, $P_{\rm ONU}$; and ii) the O-Cloud power consumption, $P_{\rm Cloud}$ \cite{Masoudi2020}.  The total power consumption is given as
\begin{equation}
    P_{\rm tot} = \sum_{l=1}^L P_{{\rm RU},l} + \sum_{l=1}^Lz_lP_{\rm ONU} + P_{\rm Cloud} \label{eq:total-power-consumption}
\end{equation}
where the binary variable $z_{l}$ indicates whether O-RU $l$ is active or not. If it is active, then $z_l=1$ and otherwise $z_l=0$. The O-RU power consumption for a particular O-RU $l$, i.e., $P_{{\rm RU},l}$ is given by \cite{Auer2011,Masoudi2020}
\begin{align} \label{eq:AP-l-energy}
    &P_{{\rm RU},l} =\nonumber\\
    & z_l \left(P_{{\rm RU},0}+\Delta^{\rm tr}\sum_{k=1}^Kx_{k,l}p_{kl}+ \mathbb{I}_s\left(P_{\rm RU,0}^{\rm proc}+ \Delta^{\rm proc}_{\rm RU} \frac{C_{{\rm RU},l}}{C_{\rm RU}^{\rm  max}}\right)\right)\nonumber \\
    =& z_l \left(P_{{\rm RU},0}+\mathbb{I}_sP_{\rm RU,0}^{\rm proc}+\Delta^{\rm proc}_{\rm RU}\frac{C_{{\rm RU},l}}{C_{\rm RU}^{\rm  max}}\right)+\Delta^{\rm tr}\sum_{k=1}^Kx_{k,l}p_{kl},
\end{align}
where $P_{{\rm RU},0}$ is the static power consumption of each O-RU when there is no transmission at the active mode and $\sum_{k=1}^Kx_{k,l}p_{kl}$ is the transmit power of O-RU $l$. The load-dependent power consumption is modeled by the slope $\Delta^{\rm tr}$. The binary indicator $\mathbb{I}_s\in\{0,1\}$ specifies which functional splitting option is used. When it is one, Option 7.2 is utilized. In this case, the low-PHY functions between the dashed and solid lines in Fig.~\ref{fig:architecture} are implemented in the processing units of the active O-RUs.  Otherwise, when $\mathbb{I}_s=0$, Option 8 is adopted and all the baseband processing is done in the O-Cloud. When $\mathbb{I}_s=1$, the term $P_{\rm RU,0}^{\rm proc}+\Delta^{\rm proc}_{\rm RU}\frac{C_{{\rm RU},l}}{C_{\rm RU}^{\rm  max}}$ in the above expression represents the processing power consumption using a load-dependent power consumption model \cite{sigwele2017energy}. $P_{\rm RU,0}^{\rm proc}$ is the idle mode processing power consumption of each active processing unit at the O-RU corresponding to zero utilization. $\Delta^{\rm proc}_{\rm RU}$ is the slope of the load-dependent processing power consumption The maximum processing capacity at each O-RU is given by $C^{\rm max}_{\rm RU}$ in giga-operations per second (GOPS) depending on the used technology. The processing utilization at O-RU $l$ is given by $0\leq C_{{\rm RU},l} \leq C^{\rm  max}_{\rm RU}$ in GOPS. Later, we will compute the required GOPS. In the second equality in \eqref{eq:AP-l-energy}, we have simplified the equation by noting that when $z_l$ is zero, $x_{k,l}$, $\forall k$ become zero by definition. Similarly, when $\mathbb{I}_s=0$, irrespective of whether O-RU $l$ is active or not, there is no processing power consumption and $C_{{\rm RU},l}$ is automatically zero.  In this case, the processing power is not included.  It is worth mentioning that if an O-RU is not active, i.e., $z_l=0$, we turn it off to save power. 

The total power consumption in the O-Cloud is computed  using the load-dependent GPP power consumption model from \cite{sigwele2017energy} as
\begin{align} \label{eq:CPU-power-consumption}
    P_{\rm Cloud} =&P_{\rm fixed}+ \frac{1}{\sigma_{\rm cool}}\Bigg( P_{\rm OLT}\sum_{w=1}^W w\ell_{w}\nonumber\\
   & +P_{{\rm GPP},0}^{\rm proc}\sum_{w=1}^Wwd_w+\Delta^{\rm proc}_{\rm GPP} \frac{C_{\rm GPP}}{C_{\rm GPP}^{\rm  max}}\Bigg),
    \end{align}
where $P_{\rm fixed}$ is the load-independent fixed power consumption that includes the power consumption of the O-Cloud dispatcher, housing facilities, etc. The cooling efficiency is $0<\sigma_{\rm cool}\leq1$. $P_{\rm OLT}$ is the power consumption of each OLT module per GPP \cite{Masoudi2020}.  The binary variable $\ell_{w}\in \{0,1\}$ is one if the LCs of $w$ number of GPPs are active and zero otherwise. Similarly, $d_w\in \{0,1\}$ is one if  $w$ number of GPPs are active for either processing or fronthaul connection to the O-RUs, and zero otherwise. This particular definition of the binary variables is to ensure that the constraints can be written as linear functions of the binary variables in the optimization problem we will consider in the next section. It is worth mentioning that the LC of an active GPP may be inactive if the corresponding GPP participates only in the processing that is redirected to it coming from other GPPs.  $P_{\rm GPP,0}^{\rm proc}$ is the idle mode processing power consumption of each active GPP in the O-Cloud corresponding to zero utilization. $\Delta^{\rm proc}_{\rm GPP}$ is the slope of the load-dependent processing power consumption of each GPP. For each GPP, the maximum processing capacity is given by $C_{\rm GPP}^{\rm max}$ in GOPS. The total processing utilization is given by $0\leq C_{{\rm GPP}} \leq W C_{\rm GPP}^{\rm  max}$ in GOPS.

\subsection{GOPS Analysis of Digital Operations at the Radio Site and O-Cloud }

In this section, we will analyze the GOPS for digital signal processing operations of a cell-free massive MIMO system.  In the uplink, if Option 8 is used, after RF processing at the O-RUs, the quantized baseband signals are directly sent to the cloud. Then, at the GPPs, cyclic prefix (CP) removal and $N_{\rm DFT}$-point DFT are performed to obtain frequency-domain signals. After resource element (RE) demapping, the remaining PHY and higher-layer operations are implemented for a particular UE. Similarly, in the downlink, after the higher-layer functions, the precoded signals are obtained. In the sequel, RE mapping, inverse DFT, and CP insertion are realized. Then, the time-domain signals are sent to the O-RUs and RF transmission takes place. If Option 7.2 is used, then the order of operations remains the same while the frequency-domain signals are sent after RE demapping to the O-Cloud in the uplink and the precoded frequency-domain signals are sent from the O-Cloud to the O-RUs for the remaining low-PHY operations in the downlink as shown in Fig.~\ref{fig:architecture}.

To compute the total GOPS in O-RU $l$, $C_{{\rm RU},l}$ and in the O-Cloud, $C_{{\rm GPP}}$, we will mainly use the results from cellular massive MIMO \cite{malkowsky2017world,desset2016massive}. In \cite{desset2016massive}, a factor two of overhead is taken in arithmetic operation calculations to account for memory operations. In the following GOPS calculations, we will also consider this by including a multiplication by two in each arithmetic operation. 
The first operation after RF processing is baseband filtering. Considering 10 taps with a polyphase filtering implementation, the corresponding complexity per O-RU is given in GOPS as $ C_{\rm filter} = 40Nf_s/10^9$ \cite{desset2016massive}. The next operation is
 DFT in the uplink and inverse DFT in the downlink, which has the complexity with fast Fourier transform (FFT) as $C_{\rm DFT} =8N N_{\rm DFT}\log_2\left(N_{\rm DFT}\right)/\left(T_{ s}10^9\right)$, which is obtained by dividing the number of required real operations by the OFDM symbol duration $T_{ s}$ \cite{malkowsky2017world}.

The GOPS of the sample-based arithmetic operations such as precoding  scale with $N_{\rm used}/T_{ s}$ \cite{malkowsky2017world}. An additional multiplying factor $\tau_d/\tau_c$ should be taken into account in precoding the downlink data since $\tau_d$ samples are precoded in each coherence block of length $\tau_c$. For the channel estimation, reciprocity calibration, and precoding computation, it scales with $N_{\rm used}/(T_{ s}\tau_c)$ since the corresponding operations are common for each sample in a coherence block of length $\tau_c$. To this end, from \cite[Sec.~6.2.2]{cell-free-book}, for LP-MMSE transmit precoding together with the required channel estimation of the served UEs by O-RU $l$ (and of the strongest UE per pilot), the GOPS (in terms of real multiplications) is computed as
\begin{align} \label{eq:complexity-PMMSE}
    &C_{{\rm prec},l} = \underbrace{\frac{N_{{\rm used}}}{T_{ s}\tau_c10^9}\left(8N\tau_p^2+8N^2\left(\tau_p+\sum_{i=1}^Kx_{i,l}\right)\right)}_{\textrm{Channel estimation}}  \nonumber\\&+ \underbrace{\frac{N_{{\rm used}}\tau_d}{T_{ s}\tau_c10^9}8N\sum_{i=1}^Kx_{i,l}}_{\textrm{Precoding}}  +  \underbrace{\frac{N_{{\rm used}}}{T_{ s}\tau_c10^9}8N\sum_{i=1}^Kx_{i,l}}_{\textrm{Reciprocity calibration}}\nonumber\\&+\underbrace{\frac{N_{{\rm used}}}{T_{ s}\tau_c10^9}\left(\left(4N^2+4N\right)\tau_p+8N^2\sum_{i=1}^Kx_{i,l}+\frac{8\left(N^3-N\right)}{3}\right)}_{\textrm{Precoding computation}} \end{align}
where we have also included the complexity of applying precoding and reciprocity calibration from \cite{malkowsky2017world}. It is worth mentioning that local precoding computation $C_{{\rm prec},l}$ in \eqref{eq:complexity-PMMSE} can be implemented at any other GPP $w$ different than the connected GPP to O-RU $l$  (benefiting from cloud sharing and virtualization via GPP dispatcher).

In addition to precoding, there are other GOPS regarding OFDM modulation/demodulation, mapping/demapping, channel coding, higher-layer control and network operations. These can be computed using the flexible power modeling in \cite{Debaillie2015a}. Let $C_{\rm other,O-RU}$ and $C_{\rm other,UE}$ denote the GOPS for the other operations which scale with the number of active O-RUs and the number of UEs that are served by each O-RU, respectively. There is also a fixed GOPS for UE operations, which are independent of the number of serving O-RUs. This is represented by $\mathcal{F}$. The total GOPS in O-RU $l$ is given by
\begin{equation} \label{eq:CRUl}
  C_{{\rm RU},l}  = \mathbb{I}_s\underbrace{\left(C_{\rm filter}+C_{\rm DFT}\right)}_{\triangleq \mathcal{S}}=\mathbb{I}_s\mathcal{S}. 
\end{equation}

If Option 8 is used, then $\mathbb{I}_s=0$ and there is no digital signal processing implemented in the O-RUs. In this case, we include the corresponding low-PHY GOPS in the total GOPS of the O-Cloud as
\begin{align} \label{eq:CGPP}
   & C_{\rm GPP}=\nonumber\\
    & \sum_{l=1}^Lz_l\left(\left(1-\mathbb{I}_s\right)\left(C_{\rm filter}+C_{\rm DFT}\right)+  C_{{\rm prec},l} +C_{\rm other,O-RU}\right) \nonumber\\&+\sum_{l=1}^L\sum_{k=1}^Kx_{k,l}C_{\rm other,UE}+\mathcal{F} \nonumber \\
    =& \mathcal{Z} \sum_{l=1}^L z_l  + \mathcal{X} \sum_{l=1}^L\sum_{k=1}^Kx_{k,l}+\mathcal{F}
\end{align}
where the constant parameters  $\mathcal{Z}$ and $\mathcal{X}$ are defined for ease of notation in the optimization problem.
 Note that when both the O-RU $l$ is active ($z_l=1$) and Option 8 is used for functional splitting ($\mathbb{I}_s=0$), the corresponding low-PHY GOPS is included in $C_{\rm GPP}$.

\section{Power-Efficient O-RU Selection, GPP and Power Allocation}\label{sec:optimization}

In this section, we will introduce the proposed optimization problem that minimizes total power consumption. The aim is to decide which O-RUs serve which UEs, i.e., the binary variables $x_{k,l}\in \{0,1\}$, the transmit powers allocated to the UEs, i.e., $p_{kl}$, which O-RUs are active and connected to the O-Cloud, i.e., $z_{l}\in \{0,1\}$, and the number of active LCs and GPPs, i.e., $\ell_w,d_w\in\{0,1\}$ in the O-Cloud. We note that the considered optimization problem is a mixed integer program since the O-RU selection together with power allocation and DU allocation  is considered, which has a combinatorial nature. To express both the objective function and the constraints in a mixed binary linear or conic form, we introduce the following additional optimization variables:
\begin{align}
&{\boldsymbol{\rho}}_k=\left [ \sqrt{p_{k1}}x_{k,1} \, \ldots \, \sqrt{p_{kL}} x_{k,L} \right ]^{\Ttran} = \left [\rho_{k,1} \, \ldots \, \rho_{k,L}\right]^{\Ttran},\\
&{\boldsymbol{\rho}}^{\prime}_l=\left [ \sqrt{p_{1l}}x_{1,l} \, \ldots \, \sqrt{p_{Kl}} x_{K,l} \right ]^{\Ttran} = \left [\rho_{1,l} \, \ldots \, \rho_{K,l}\right]^{\Ttran}.
 \end{align}

Due to the limited capacity of each wavelength in TWDM-PON, which is denoted by $R_{\rm max}$, we can assign at most
\begin{equation}
W_{\rm max} = \left \lfloor \frac{R_{\rm max}}{\mathbb{I}_sR_{\rm fronthaul}^{(7.2)}+(1-\mathbb{I}_s)R_{\rm fronthaul}^{(8)}} \right \rfloor  
\end{equation}
O-RUs to each wavelength and,
hence, to each GPP $w$, for $w = 1, \ldots , W$.
 When $\mathbb{I}_s=1$, the functional split option 7.2 is used with the corresponding required fronthaul data rate $R_{\rm fronthaul}^{(7.2)}$. On the other hand, when $\mathbb{I}_s=0$, the functional split option 8 is used with the required fronthaul data rate $R_{\rm fronthaul}^{(8)}$.   

In the considered network power consumption minimization problem, we assume each UE $k$ has a SE request with the corresponding minimum SINR requirement $\gamma_k$. Hence, we have QoS constraints in the form of $\mathacr{SINR}_{k}\geq \gamma_k$ for each UE $k$. The optimization problem can be cast using the introduced optimization variables as
\begin{subequations} \label{eq:optimization1}
\begin{align} 
&\underset{\begin{subarray}{c}
  z_l, x_{k,l}, \forall k, \forall l \\
  \ell_w, d_w, \forall w \\
  \boldsymbol{\rho}_k, \forall k
  \end{subarray} }{\textrm{minimize}} \quad  P_{\rm fixed} +\left(P_{{\rm RU},0} +P_{\rm ONU}\right)\sum_{l=1}^Lz_{l}\nonumber \\ &+  \left(\mathbb{I}_s\left(P_{\rm RU,0}^{\rm proc}+ \frac{\Delta^{\rm proc}_{\rm RU}\mathcal{S}} {C_{\rm RU}^{\rm  max}}\right)+\frac{\Delta^{\rm proc}_{\rm GPP}\mathcal{Z}}{ \sigma_{\rm cool}C_{\rm GPP}^{\rm  max}}\right)\sum_{l=1}^Lz_{l} \nonumber\\
  &+\Delta^{\rm tr}\sum_{l=1}^L\sum_{k=1}^K\rho_{k,l}^2+ \frac{P_{\rm OLT}}{\sigma_{\rm cool}}\sum_{w=1}^W w\ell_{w}+\frac{P_{\rm GPP,0}^{\rm proc}}{\sigma_{\rm cool}}\sum_{w=1}^Wwd_w \nonumber\\&
  +\frac{\Delta^{\rm proc}_{\rm GPP}\mathcal{X}}{\sigma_{\rm cool}C_{\rm GPP}^{\rm max}}\sum_{l=1}^L\sum_{k=1}^K x_{k,l}+\frac{\Delta^{\rm proc}_{\rm GPP}\mathcal{F}}{\sigma_{\rm cool}C_{\rm GPP}^{\rm max}} \label{eq:Xobjective} \\
&\textrm{subject to:} \nonumber \\
&\frac{\left|\vect{b}_k^{\Ttran}{\boldsymbol{\rho}}_k\right|^2}{\sum\limits_{i=1}^K{\boldsymbol{\rho}}_i^{\Ttran}{\vect{C}}_{ki}{\boldsymbol{\rho}}_i+\sigma^2}\geq \gamma_k, \quad \forall k, \label{eq:Xconstraint1} \\
&  \sum_{l=1}^Lz_l \leq W_{\rm max}W, \label{eq:Xconstraint2} \\
&  \frac{\sum_{k=1}^Kx_{k,l}}{K} \leq z_{l} \leq \sum_{k=1}^Kx_{k,l},  \quad \forall l, \label{eq:Xconstraint4} \\
& \hspace{0mm} \sum_{w=1}^{W}(w-1)\ell_{w}\leq \frac{\sum_{l=1}^Lz_l }{W_{\rm max}} \leq  \sum_{w=1}^{W}w\ell_w, \label{eq:Xconstraint5} \\
& \mathcal{Z} \sum_{l=1}^L z_l  + \mathcal{X} \sum_{l=1}^L\sum_{k=1}^Kx_{k,l}+\mathcal{F}\leq C_{\rm GPP}^{\rm  max} \sum_{w=1}^W wd_w, \label{eq:Xconstraint6}  \\
&  \sum_{w=1}^W \ell_w= 1, \quad \sum_{w=1}^W d_{w} = 1, \label{eq:Xconstraint8} \\
&  \sum_{w=1}^W w\ell_w \leq \sum_{w=1}^W wd_w,  \label{eq:Xconstraint9}\\
& 0\leq\rho_{k,l}\leq \sqrt{p_{\rm max}}x_{k,l}, \quad \forall k, \forall l, \label{eq:Xconstraint11} \\
& \left\Vert {\boldsymbol{\rho}}^{\prime}_l \right\Vert\leq \sqrt{p_{\rm max}}z_l, \quad  \forall l, \label{eq:Xconstraint12} \\
& z_l,   \ell_w, d_w, x_{k,l} \in \{0,1\}, \quad \forall k, \forall l, \forall w. \label{eq:Xconstraint13}
\end{align}
\end{subequations}
The constraints in \eqref{eq:Xconstraint1} are to guarantee that each UE's minimum SINR requirement is satisfied. The constraint in \eqref{eq:Xconstraint2} guarantees that the number of active O-RUs is determined  not to exceed the maximum allowable number determined by the fronthaul limitations of the given functional splitting option. The constraints in \eqref{eq:Xconstraint4} relate the binary variables $x_{k,l}$ and $z_l$, i.e., an O-RU is active if and only if it serves at least one UE. The constraint in \eqref{eq:Xconstraint5} connects the number of active O-RUs  to the the number of required active LCs. The constraint in \eqref{eq:Xconstraint6} guarantees that the total GOPS does not exceed the processing capability of active GPPs in the O-Cloud. The constraints in \eqref{eq:Xconstraint8} are to satisfy that $\ell_w$ and $d_w$ are only one for one value of $w$ since these binary variables are one when the number of active LCs or GPPs is equal to their sub-index. The constraint in \eqref{eq:Xconstraint9} is to ensure that the number of active GPPs is always greater than or equal to the number of active LCs. The constraints in \eqref{eq:Xconstraint11} guarantee that the square root of the power coefficient for UE $k$ and O-RU $l$ is zero if UE $k$ is not served by O-RU $l$. Here, $p_{\rm max}$ is the maximum transmit power budget of each O-RU and when $x_{k,l}=1$, this constraint does not limit $\rho_{k,l}$. \eqref{eq:Xconstraint12} represents the per-O-RU transmit power constraints. Finally, the constraints in \eqref{eq:Xconstraint13} specify which variables are binary.

Note that the SINR constraints in \eqref{eq:Xconstraint1} can be re-written in second-order cone form \cite[Sec.~7.1.2]{cell-free-book}. As a result, the overall optimization problem is a mixed binary second-order cone programming problem, which has a convex structure except for the binary constraints. Hence, the global optimum can be obtained by the branch-and-bound algorithm \cite{gurobi}. It is known that the complexity grows exponentially with the number of discrete variables, which in our case means the number of O-RUs, UEs, and GPPs.  In the next part, we will develop a lower complexity solution by approximating the problem and using concave programming.

\subsection{Approximate Optimization Problem Formulation Using $l_0$ Norm Minimization }

In devising a lower complexity algorithm, the main step is to eliminate the binary variables in the optimization problem. To this end, we will keep the SINR constraints and some other few constraints as they are while reflecting the other constraints in the objective function by using novel approximations that allow elimination of several binary variables and obtaining a concave objective function with convex constraints at the end. Let us go through all the constraints in \eqref{eq:Xconstraint1}-\eqref{eq:Xconstraint13} in the sequel. First, we express the SINR constraints in \eqref{eq:Xconstraint1} in second-order cone form as 
\begin{equation}
\left \| \begin{bmatrix} \sqrt{\gamma_k}\vect{C}_{k1}^{\frac{1}{2}}\boldsymbol{\rho}_1 \\ \vdots \\ \sqrt{\gamma_k}\vect{C}_{kK}^{\frac{1}{2}}\boldsymbol{\rho}_K \\
\sqrt{\gamma_k}\sigma
\end{bmatrix} \right\|  \leq  \vect{b}_k^{\Ttran}{\boldsymbol{\rho}}_k,   \quad \forall k. \label{eq:Xconstraint1b}
\end{equation}
To simplify the problem and make it manageable, we will not consider the constraints in \eqref{eq:Xconstraint2}-\eqref{eq:Xconstraint9} and the binary constraints in \eqref{eq:Xconstraint13}. Hence, the variables  $\ell_w$ and $d_w$ do not exist in the modified problem. To reflect their impact on the objective function, we will approximate the values of   $\sum_{w=1}^Ww\ell_w$ and $\sum_{w=1}^Wwd_w$ in the objective function \eqref{eq:Xobjective} using the constraints in  \eqref{eq:Xconstraint5}-\eqref{eq:Xconstraint9} as follows:
\begin{align}
&  \sum_{w=1}^{W}w\ell_{w}\approx \frac{\sum_{l=1}^Lz_l }{W_{\rm max}} = \frac{\|\vect{z}\|_0}{W_{\rm max}}, \label{eq:Xconstraint5b} \\
& \sum_{w=1}^W wd_w \approx \max \left(  \sum_{w=1}^{W}w\ell_{w}, \frac{  \mathcal{Z} \|\vect{z}\|_0  +\mathcal{X}\sum_{k=1}^K \|\boldsymbol{\rho}_k\|_0+\mathcal{F}}{C_{\rm GPP}^{\rm  max}}\right) \label{eq:Xconstraint6b} 
\end{align}
where $ \vect{z}=[z_1 \ \ldots \ z_L]^{\Ttran}$
and we have used the fact that $\sum_{l=1}^Lx_{k,l}=\|\boldsymbol{\rho}_k\|_0$. To write $\sum_{w=1}^W wd_w$ in the objective function of the approximate modified problem, we can eliminate the maximum operation and approximate $\sum_{w=1}^W wd_w$ by using the upper bound of  \eqref{eq:Xconstraint6b} as
\begin{align}
 \sum_{w=1}^W wd_w \approx &  \sum_{w=1}^{W}w\ell_{w}+ \frac{  \mathcal{Z} \|\vect{z}\|_0  +\mathcal{X}\sum_{k=1}^K \|\boldsymbol{\rho}_k\|_0+\mathcal{F}}{C_{\rm GPP}^{\rm  max}} \nonumber \\
\approx & \frac{\|\vect{z}\|_0}{W_{\rm max}} + \frac{  \mathcal{Z} \|\vect{z}\|_0  +\mathcal{X}\sum_{k=1}^K \|\boldsymbol{\rho}_k\|_0+\mathcal{F}}{C_{\rm GPP}^{\rm  max}}. \label{eq:Xconstraint6c} 
\end{align}

Keeping the constraints  \eqref{eq:Xconstraint12} and the left-side constraint in \eqref{eq:Xconstraint11}, and neglecting the fixed part of the objective function, the approximated optimization problem in terms of reduced number of variables can be expressed as
\begin{subequations}
 \label{eq:optimization2}
\begin{align} 
& \underset{
 \vect{z}, 
  \boldsymbol{\rho}_k, \forall k
   }{\textrm{minimize}} \quad     \left(P_{{\rm RU},0} +P_{\rm ONU}\right)\|\vect{z}\|_0 \nonumber\\
   &+\left(\mathbb{I}_s\left(P_{\rm RU,0}^{\rm proc}+ \frac{\Delta^{\rm proc}_{\rm RU}\mathcal{S}}{C_{\rm RU}^{\rm  max}}\right)+\frac{\Delta^{\rm proc}_{\rm GPP}\mathcal{Z}}{ \sigma_{\rm cool}C_{\rm GPP}^{\rm  max}}\right)\|\vect{z}\|_0\nonumber\\
   &+\Delta^{\rm tr}\sum_{l=1}^L\sum_{k=1}^K\rho_{k,l}^2+ \frac{P_{\rm OLT}+P_{\rm GPP,0}^{\rm proc}}{\sigma_{\rm cool}}\frac{\|\vect{z}\|_0}{W_{\rm max}}\nonumber\\
&+\frac{P_{\rm GPP,0}^{\rm proc}}{\sigma_{\rm cool}C_{\rm GPP}^{\rm max}}\left( \mathcal{Z} \|\vect{z}\|_0  +\mathcal{X}\sum_{k=1}^K \|\boldsymbol{\rho}_k\|_0\right)\nonumber\\
&+\frac{\Delta^{\rm proc}_{\rm GPP}\mathcal{X}}{\sigma_{\rm cool}C_{\rm GPP}^{\rm max}}\sum_{k=1}^K \|\boldsymbol{\rho}_k\|_0 \label{eq:Yobjective} \\
&\textrm{subject to:} \nonumber \\
&\left \| \begin{bmatrix} \sqrt{\gamma_k}\vect{C}_{k1}^{\frac{1}{2}}\boldsymbol{\rho}_1 \\ \vdots \\ \sqrt{\gamma_k}\vect{C}_{kK}^{\frac{1}{2}}\boldsymbol{\rho}_K \\
\sqrt{\gamma_k}\sigma
\end{bmatrix} \right\|  \leq  \vect{b}_k^{\Ttran}{\boldsymbol{\rho}}_k,   \quad \forall k, \label{eq:Yconstraint1} \\
&    \left\Vert {\boldsymbol{\rho}}^{\prime}_l \right\Vert\leq \sqrt{p_{\rm max}}z_l,  \quad  \forall l \\
&\left\Vert {\boldsymbol{\rho}}^{\prime}_l \right\Vert\leq \sqrt{p_{\rm max}},    \, \, \forall l, \quad \rho_{k,l}\geq 0,  \, \, \forall k, \forall l \label{eq:Yconstraint2}
\end{align}
\end{subequations}
where we have included $\left\Vert {\boldsymbol{\rho}}^{\prime}_l \right\Vert\leq \sqrt{p_{\rm max}}$ in \eqref{eq:Yconstraint2} to guarantee the per-O-RU power constraints are satisfied even if $z_l>1$ for some $l$.

Note that $l_0$ norm is not a real norm and $\Vert.\Vert_0$ is not a convex function. Hence, the optimization problem in \eqref{eq:optimization2} is not convex. One way to solve this problem in an efficient way is to replace the $l_0$ norm by a more tractable function.  After evaluating several methods, we have observed that the following approach leads to decent performance with nice convergence properties in the considered iterative algorithm. The following continuously differentiable concave function can be used in place of $\Vert \vect{z}\Vert_0$:
\begin{equation} \label{eq:concave-approx}
    f(\vect{z}) =  \sum_{l=1}^L(1-e^{-\alpha z_l})
\end{equation}
with $\alpha>0$ \cite{rinaldi2010concave}. It can be seen that as $\alpha \to \infty$, $f(\vect{z})\to \Vert \vect{z}\Vert_0$. Using the concave function  $f(\vect{z})$, the non-convex problem in \eqref{eq:optimization2} can be approximated as
\begin{subequations}
 \label{eq:optimization3}
\begin{align} 
&\underset{
 \vect{z}, 
  \boldsymbol{\rho}_k, \forall k
   }{\textrm{minimize}} \quad   \mathcal{A}_{z}f(\vect{z}) +\mathcal{A}_{\rho}\sum_{k=1}^Kf(\boldsymbol{\rho}_k) +\Delta^{\rm tr}\sum_{l=1}^L\sum_{k=1}^K\rho_{k,l}^2    \label{eq:Zobjective} \\
&\textrm{subject to} \quad \eqref{eq:Yconstraint1}-\eqref{eq:Yconstraint2}
\end{align}
\end{subequations}
where the constants $\mathcal{A}_{i}$, for $i\in\{z,\rho\}$ in the objective function are obtained by summing the terms that multiply the corresponding $\Vert \cdot\Vert_0$ terms in  \eqref{eq:Yobjective}. The above problem
is in concave programming form. It is not convex, but the so-called  \emph{concave-convex procedure (CCP)} outlined in Algorithm~\ref{alg:ccp} can be used, where the concave objective function is convexified around the previous solution and a convex problem is solved at each iteration. From \cite[Thm.~4]{sriperumbudur2009convergence}, Algorithm~\ref{alg:ccp} converges to a stationary point of the problem in \eqref{eq:optimization3} under suitable constraint qualification when \eqref{eq:optimization3} is feasible.

The solution found by Algorithm~\ref{alg:ccp} can further be improved by enforcing more sparsity, and thus less number of active O-RUs, via the refinement steps outlined in Algorithm~\ref{alg:ccp2}. First, the power coefficients $\rho_{k,l}$, whose values normalized by the maximum power coefficient are smaller than the threshold $0<\zeta\ll 1$, are set to zero. Then, the number of active O-RUs is determined by checking the non-zero power coefficients. Next, the problem of minimizing radio site power consumption under the SINR and per-O-RU power constraints given in \eqref{eq:optimization4-b} is solved. The number of active O-RUs is iteratively reduced according to the O-RU transmit powers found previously until this problem is infeasible. Once infeasibility is detected, the lastly found power coefficients are returned as the output of Algorithm~\ref{alg:ccp2}. Here, the threshold $\zeta$ should be sufficiently small to ensure that the problem in \eqref{eq:optimization4-b} is feasible at the first iteration, but also fine-tuned in a way to eliminate unnecessarily small power coefficients that do not affect the feasibility of the power allocation problem.

\begin{figure}[t]
\begin{algorithm}[H]
	\caption{CCP algorithm for solving the problem in \eqref{eq:optimization3}.} \label{alg:ccp}
	\begin{algorithmic}[1]
		\State {\bf Initialization:} 
		\begin{itemize}
			\item Set $\vect{z}^{(0)}$ and $\boldsymbol{\rho}_k^{(0)}$, for $k=1,\ldots,K$, with arbitrary positive entries and the solution accuracy $\varepsilon>0$.
			\item Set the iteration counter $t=0$. 
		\end{itemize}
		\Repeat
		\State  $t \gets t+1$.
	    	\State Solve the convex problem
  \begin{subequations}
 \label{eq:optimization4}
\begin{align} 
\underset{
 \vect{z}, 
  \boldsymbol{\rho}_k, \forall k
   }{\textrm{minimize}} \quad   &\Delta^{\rm tr}\sum_{l=1}^L\sum_{k=1}^K\rho_{k,l}^2+\mathcal{A}_{z}\nabla f\left(\vect{z}^{(t-1)}\right)^{\Ttran}\vect{z} \nonumber\\
   &+ \mathcal{A}_{\rho}\sum_{k=1}^K\nabla f\left(\boldsymbol{\rho}_k^{(t-1)}\right)^{\Ttran}\boldsymbol{\rho}_k     \label{eq:Tobjective} \\
\textrm{subject to} \quad &\eqref{eq:Yconstraint1}-\eqref{eq:Yconstraint2}.
\end{align}
\end{subequations}
\State Set $\vect{z}^{(t)}$,  $\boldsymbol{\rho}_1^{(t)},\ldots,\boldsymbol{\rho}_K^{(t)}$ as the solution of \eqref{eq:optimization4}.
	   	\Until{the normalized squared error difference between the current and previous objective functions in \eqref{eq:Zobjective} is less than $\varepsilon$.}
	\State	{\bf Output:} $\vect{z}^{(t)}$,  $\boldsymbol{\rho}_1^{(t)},\ldots,\boldsymbol{\rho}_K^{(t)}$.
	\end{algorithmic}
\end{algorithm}
\end{figure}

	\begin{figure}[t]
\begin{algorithm}[H]
	\caption{Refinement algorithm for improving the solution found in Algorithm~\ref{alg:ccp}.} \label{alg:ccp2}
	 \begin{algorithmic}[1]
		\State {\bf Initialization:} \begin{itemize}
		    \item Set  $\boldsymbol{\rho}_k^{(0)}$, for $k=1,\ldots,K$, as the output of Algorithm~\ref{alg:ccp}.
      \item Compute $\overline{\rho}= \max\limits_{l=1,\ldots,L, k=1,\ldots,K} \rho_{k,l}^{(0)}$. 
      \item Set the power coefficients $\rho_{k,l}^{(0)}$, which are sufficiently small so that $\rho_{k,l}^{(0)}/\overline{\rho}\leq \zeta$, to zero, where $0<\zeta\ll 1$ is the threshold parameter.
      \item Determine the number of active O-RUs, denoted by $L_{\rm active}^{(0)}$, as the number of O-RUs $l$ so that 
  \begin{align}
      \sum_{k=1}^K\left(\rho_{k,l}^{(0)}\right)^2>0.
\end{align}
\item Set the iteration counter $t=0$.
\end{itemize}
		\Repeat
		\State  $t \gets t+1$.
	    	\State Solve the convex problem	
  \begin{subequations}
 \label{eq:optimization4-b}
\begin{align} 
&\underset{ 
  \boldsymbol{\rho}_k, \forall k
   }{\textrm{minimize}} \quad   \Delta^{\rm tr}\sum_{l=1}^L\sum_{k=1}^K\rho_{k,l}^2 
 \label{eq:Tobjective-b} \\
&\textrm{subject to:} \nonumber \\
&\eqref{eq:Yconstraint1} \\
&\left\Vert {\boldsymbol{\rho}}^{\prime}_l \right\Vert\leq \sqrt{p_{\rm max}},    \quad \forall l, \quad \rho_{k,l}\geq 0,  \quad \forall k, \quad \forall l \\
& \rho_{k,l}=0, \quad \textrm{for } \rho_{k,l}^{(t-1)}=0, \quad \forall k, \quad \forall l.
\end{align}
\end{subequations}
\State  \begin{itemize}
	\item If the problem is feasible, set  $\boldsymbol{\rho}_1^{(t)},\ldots,\boldsymbol{\rho}_K^{(t)}$ as the solution of \eqref{eq:optimization4-b}. 
	\item Reduce the number of active O-RUs by one, i.e.,  $L_{\rm active}^{(t)}=L_{\rm active}^{(t-1)}-1$.
	\item Set $\rho_{k,l}^{(t)}$ to zero for the $L-L_{\rm active}^{(t)}$ O-RUs that have the least total transmit powers $\sum_{k=1}^K\left(\rho_{k,l}^{(t)}\right)^2$.
	\end{itemize}
	   	\Until{An infeasible solution is obtained or $L_{\rm active}^{(t)}=1$.}
	\State	{\bf Output:}  $\boldsymbol{\rho}_1^{(t)},\ldots,\boldsymbol{\rho}_K^{(t)}$.
	\end{algorithmic}
\end{algorithm}
\end{figure}

\section{ Joint Sum-SE Maximization  and Power Consumption Minimization}\label{sec:sumrate}

In this section, we will consider a multiobjective optimization problem that aims to jointly minimize the total end-to-end power consumption and maximize the sum-SE of all the UEs. To this end, we remove the SINR constraints in \eqref{eq:Yconstraint1} from the problem \eqref{eq:optimization3} and include the sum-SE with a certain weight to the objective function. The considered problem can be cast as
\begin{subequations}
 \label{eq:sumSE-optimization}
\begin{align} &
\underset{
 \vect{z}, 
  \boldsymbol{\rho}_k, \forall k
   }{\textrm{minimize}} \quad \Delta^{\rm tr}\sum_{l=1}^L\sum_{k=1}^K\rho_{k,l}^2 +  \mathcal{A}_{z}f(\vect{z}) +\mathcal{A}_{\rho}\sum_{k=1}^Kf(\boldsymbol{\rho}_k) \nonumber\\& -\lambda \sum_{k=1}^K \ln \left(1+\frac{\left|\vect{b}_k^{\Ttran}{\boldsymbol{\rho}}_k\right|^2}{\sum\limits_{i=1}^K{\boldsymbol{\rho}}_i^{\Ttran}{\vect{C}}_{ki}{\boldsymbol{\rho}}_i+\sigma^2}\right)    \label{eq:sumSE-objective} \\
&\textrm{subject to:} \nonumber\\
& \left\Vert {\boldsymbol{\rho}}^{\prime}_l \right\Vert\leq \sqrt{p_{\rm max}}z_l,  \quad  \forall l \\
&\left\Vert {\boldsymbol{\rho}}^{\prime}_l \right\Vert\leq \sqrt{p_{\rm max}},    \, \, \forall l, \quad \rho_{k,l}\geq 0,  \, \, \forall k, \forall l \label{eq:sumSE-constraint}
\end{align}
\end{subequations}
where the parameter $\lambda>0$ determines the weighting of the sum-SE term in the multi-objective function. We have used $\ln(\cdot)$ instead of $\log_2(\cdot)$ for simplicity without loss of generality since only a constant factor differs between them. Moreover, the pre-log factor in  \eqref{eq:downlink-rate-expression-level2} is not considered since its effect can be absorbed into the penalty parameter $\lambda$. Weighted MMSE is a common technique to find local optima of sum-SE maximization problems under power constraints \cite{Shi2011}. However, here our problem differs from the traditional sum-SE maximization structure since it also includes the total power consumption in the objective function, which is in a non-convex form. In this paper, we will devise an alternative algorithm to take advantage of CCP that we have already constructed for the power minimization problem in the previous section. We first utilize the transform of $\ln(1+\mathacr{SINR}_k)$ to an equivalent form from  \cite[Thm.~3]{shen2018fractional2}. Using the newly defined optimization variables $\chi_k$, for $k=1,\ldots,K$, the optimization problem in \eqref{eq:sumSE-optimization} can be equivalently (in terms of the optimal solution) expressed as
\begin{subequations}
 \label{eq:sumSE-optimization2}
\begin{align} 
&\underset{
 \vect{z}, 
  \boldsymbol{\rho}_k, \chi_k,\forall k
   }{\textrm{minimize}} \quad  \Delta^{\rm tr}\sum_{l=1}^L\sum_{k=1}^K\rho_{k,l}^2+ \mathcal{A}_{z}f(\vect{z}) +\mathcal{A}_{\rho}\sum_{k=1}^Kf(\boldsymbol{\rho}_k) \nonumber \\
   &-\lambda \left( \sum_{k=1}^K \ln(1+\chi_k)-\sum_{k=1}^K\chi_k\right)\nonumber\\
   &-\lambda\sum_{k=1}^K(1+\chi_k) \frac{\left|\vect{b}_k^{\Ttran}{\boldsymbol{\rho}}_k\right|^2}{\sum\limits_{i=1}^K{\boldsymbol{\rho}}_i^{\Ttran}{\vect{C}}_{ki}{\boldsymbol{\rho}}_i+\left|\vect{b}_k^{\Ttran}{\boldsymbol{\rho}}_k\right|^2+\sigma^2}    \label{eq:sumSE-objective2} \\
&\textrm{subject to:} \nonumber\\
&\left\Vert {\boldsymbol{\rho}}^{\prime}_l \right\Vert\leq \sqrt{p_{\rm max}}z_l,  \quad \quad \forall l \\
&\left\Vert {\boldsymbol{\rho}}^{\prime}_l \right\Vert\leq \sqrt{p_{\rm max}},     \, \, \forall l, \quad \rho_{k,l}\geq 0,  \, \, \forall k, \forall l \label{eq:sumSE-constraint2}
\end{align}
\end{subequations}
where the equivalency can be shown by taking the derivative of the objective function with respect to $\chi_k$, for $k=1,\ldots,K$, equating them to zero, and inserting the optimal $\chi_k$ to the objective function. Next, we will develop a further novel transformation and integrate CCP to solve the resulting problem. The first step is to introduce the optimization variables $u_k$ and $r_k$, for $k=1,\ldots,K$, to represent upper bounds to $(1+\chi_k)^{-1}$ and  $\left(\sum\limits_{i=1}^K{\boldsymbol{\rho}}_i^{\Ttran}{\vect{C}}_{ki}{\boldsymbol{\rho}}_i+\left|\vect{b}_k^{\Ttran}{\boldsymbol{\rho}}_k\right|^2+\sigma^2\right)/(1+\chi_k)$ in \eqref{eq:sumSE-objective2}, respectively. We then re-cast the problem in   \eqref{eq:sumSE-optimization2} as 
\begin{subequations}
 \label{eq:sumSE-optimization3}
\begin{align} &
\underset{
 \vect{z}, 
  \boldsymbol{\rho}_k, \chi_k, u_k,r_k,\forall k
   }{\textrm{minimize}} \   \Delta^{\rm tr}\sum_{l=1}^L\sum_{k=1}^K\rho_{k,l}^2+\mathcal{A}_{z}f(\vect{z}) +\mathcal{A}_{\rho}\sum_{k=1}^Kf(\boldsymbol{\rho}_k) \nonumber\\
   &+\lambda \left( \sum_{k=1}^K \ln(u_k)+\sum_{k=1}^K\chi_k-\sum_{k=1}^K \frac{\left(\vect{b}_k^{\Ttran}{\boldsymbol{\rho}}_k\right)^2}{r_k }\right)    \label{eq:sumSE-objective3} \\
&\textrm{subject to:} \nonumber\\
&\left\Vert {\boldsymbol{\rho}}^{\prime}_l \right\Vert\leq \sqrt{p_{\rm max}}z_l,  \quad \forall l \\
&\left\Vert {\boldsymbol{\rho}}^{\prime}_l \right\Vert\leq \sqrt{p_{\rm max}},    \, \, \forall l, \quad \rho_{k,l}\geq 0,  \, \, \forall k, \forall l  \label{eq:sumSE-constraint3} \\
& \left \| \begin{bmatrix} \sqrt{2}\vect{C}_{k1}^{\frac{1}{2}}\boldsymbol{\rho}_1 \\ \vdots \\ \sqrt{2}\vect{C}_{kK}^{\frac{1}{2}}\boldsymbol{\rho}_K \\ \sqrt{2}\vect{b}_k^{\Ttran}{\boldsymbol{\rho}}_k \\
\sqrt{2}\sigma \\ (1+\chi_k) \\ r_k
\end{bmatrix}^{\Ttran} \right\|  \leq  1+\chi_k+r_k, \quad \forall k \label{eq:sumSE-constraint3b} \\
& \left \| \begin{bmatrix} (1+\chi_k)  \\ u_k \\ \sqrt{2}
\end{bmatrix} \right\|  \leq  1+\chi_k+u_k,   \quad \forall k\label{eq:sumSE-constraint3c} 
\end{align}
\end{subequations}
where we have used second-order cone constraints to construct the inequalities $(1+\chi_k)^{-1}\leq u_k$ and $\left(\sum\limits_{i=1}^K{\boldsymbol{\rho}}_i^{\Ttran}{\vect{C}}_{ki}{\boldsymbol{\rho}}_i+\left|\vect{b}_k^{\Ttran}{\boldsymbol{\rho}}_k\right|^2+\sigma^2\right)/(1+\chi_k)\leq r_k$. The following lemma demonstrates the equivalency between the problems  \eqref{eq:sumSE-optimization2} and  \eqref{eq:sumSE-optimization3} in terms of the optimal solutions.

\begin{lemma}
The optimal values of $\{\vect{z},
\boldsymbol{\rho}_k, \forall k\}$ are the same for the problems \eqref{eq:sumSE-optimization2} and   \eqref{eq:sumSE-optimization3}.

\begin{proof}To prove the claim, it is enough to show that the constraints in \eqref{eq:sumSE-constraint3b}-\eqref{eq:sumSE-constraint3c} are satisfied with equality at the optimal solution. We can prove this by contradiction. Assume that for the optimal solution of \eqref{eq:sumSE-optimization3}, at least one of the constraints in \eqref{eq:sumSE-constraint3b} is satisfied with strict inequality. Then, we can reduce the value of respective $r_k$ until the corresponding constraint is satisfied with equality without violating any other constraint. In this way, we also improve the objective value, which contradicts that the initial $r_k$ is optimum. Hence, all the constraints in \eqref{eq:sumSE-constraint3b} are satisfied with equality. Moreover, assume that at least one of the constraints in \eqref{eq:sumSE-constraint3c} is satisfied with strict inequality, which leads to $(1+\chi_k)^{-1}<u_k$, for the corresponding $k$. Then, we can reduce the value of $u_k$ until $(1+\chi_k)^{-1}=u_k$ without violating any constraints, and improving the objective function. This contradicts that $(1+\chi_k)^{-1}<u_k$ for at least one $k$. Hence, all the constraints in \eqref{eq:sumSE-constraint3c} are satisfied with equality.  Then, by inserting the value of $r_k$ and $(1+\chi_k)^{-1}=u_k$ into the objective function, the problems  \eqref{eq:sumSE-optimization2} and   \eqref{eq:sumSE-optimization3} can be shown in identical form.
\end{proof}
\end{lemma}

We note that the objective function in \eqref{eq:sumSE-objective3} is a summation of a  convex and concave  function explained as follows. The first part of the objective function, i.e., $\Delta^{\rm tr}\sum_{l=1}^L\sum_{k=1}^K\rho_{k,l}^2$ is a convex function. The term $\mathcal{A}_{z}f(\vect{z}) +\mathcal{A}_{\rho}\sum_{k=1}^Kf(\boldsymbol{\rho}_k)$ is a concave function. Let's go through the remaining terms one by one. We note that $\ln(u_k)$ is a concave function, which can be shown by the sign of its second derivative. The last term $-\frac{\left(\vect{b}_k^{\Ttran}{\boldsymbol{\rho}}_k\right)^2}{r_k }$ is a concave function of $\boldsymbol{\rho}_k$ and $r_k$ since $g(t_k,r_k)=\frac{t_k^2}{r_k}$, which is a quadratic-over-linear function, is convex in terms of $t_k, r_k$, where $t_k=\vect{b}_k^{\Ttran}{\boldsymbol{\rho}}_k$. We also note that the constraints in  \eqref{eq:sumSE-constraint3b}-\eqref{eq:sumSE-constraint3c} are convex, and, thus, we can use CCP by solving a  convexified problem at each iteration. Furthermore, the convexified problem has a quadratic objective function with second-order cone constraints, which enables the use of efficient algorithms thanks to our unique reformulation. The steps of the respective algorithm are similar to those of Algorithm~\ref{alg:ccp}, so we do not repeat it for brevity.

\subsection{End-to-end vs. local coordination-based vs. radio-only resource orchestration}\label{sec:methods}

The optimization problems that have been considered so far provide the user-centric O-RU clusters, active O-RUs, and the respective transmit power allocation. Based on that, the processing requirements in terms of GOPS and the required number of active LCs (TWDM-PON wavelengths) and the number of GPPs are computed for three different resource orchestration schemes. 

(i) Fully virtualized end-to-end resource orchestration: The fronthaul resources and connections between the O-RUs, LCs, and GPPs are assumed to be fully virtualized and jointly orchestrated so that the minimal number of LCs and GPPs are activated. Number of LCs is set as the smallest integer that is greater than or equal to the number of active O-RUs divided by the number of O-RUs that each LC can serve. The number of active GPPs, which affect the GPP idle power, is also determined as the smallest possible value in the O-Cloud.

(ii) Local coordination-based resource orchestration: This is not a fully virtualized system, where the fronthaul resources for each O-RU are fixed. The number of active LCs is determined based on the dedicated optical wavelengths to the active O-RUs. Although full virtualization is assumed for UE-specific operations in the GPPs, the O-RU-specific operations are assumed to be executed in the dedicated GPPs to the fixed optical LC and wavelength serving a particular O-RU. The respective processing resources are virtualized among the GPPs allocated to the O-RUs that share the same optical wavelength. Hence, an increase in the fronthaul and processing idle power consumption is expected in the local-coordination-based resource orchestration. In such a system, the only power-saving mechanism is shutting down unused O-RUs and LCs, and partial virtualization within GPPs connected to each LC.

(iii) Radio-only scheme: The fronthaul and cloud resources are fixed. Although the load-dependent and O-RU powers facilitate the same sparsity induced by the proposed algorithms, the number of active LCs is selected according to the fixed fronthaul connections, and the number of active GPPs is selected under the peak traffic assumption in which all O-RUs are active. The only power-saving mechanism is shutting down unused O-RUs and LCs, which explains why we call this scheme ``radio-only'' resource allocation. 

It is worth pointing out that in all three resource allocation schemes, the GPP and O-RU load-dependent power consumption is the same, where the difference lies in the GPP idle power and fronthaul power consumption. The second and third schemes are developed as benchmarks.

\section{Numerical Results and Discussion}\label{sec:numerical}

In this part, we quantify the end-to-end power consumption of a cell-free massive MIMO system in the O-RAN architecture to gain an understanding of the impact of joint radio, fronthaul, and cloud resource allocation, the selection of the functional splitting option, and the most power-consuming network components. Three resource orchestration schemes, which are described in Section~\ref{sec:sumrate}.A, are considered: i) fully virtualized end-to-end; ii) local coordination-based; and iii) radio-only resource allocation. 

The simulation parameters are outlined in Table~\ref{tab:simulation}, and they are mainly set from the works \cite{Demir2022ICC,malkowsky2017world,Auer2011,wang2016energy,Masoudi2020,sigwele2017energy,simeonidou2020dynamic}.
In particular, we consider pico-cell power parameters from \cite{Auer2011}. The GOPS/Watt for each of $W$ GPPs in the cloud and the processing unit of each O-RU is 2.434 according to 2x Intel Xeon E5-2683 v4 processor from \cite[Tab.~1]{simeonidou2020dynamic}.  The idle power $P_{\rm GPP,0}^{\rm proc}=P_{\rm RU,0}^{\rm proc}$ and the slope $\Delta^{\rm proc}_{\rm GPP}=\Delta^{\rm proc}_{\rm RU}$ are scaled linearly such that each GPP in the cloud and O-RU processing unit has $C_{\rm GPP}^{ \rm max}=C_{\rm RU}^{ \rm max}=180$\,GOPS as in \cite{sigwele2017energy}. The deployment and radio site parameters are as in the running example of \cite[Sec.~5.3]{cell-free-book} with $f_c=2$\,GHz except for the parameters that are listed in Table~\ref{tab:simulation}. The functional split options 8 and 7.2 are denoted as FS-8 and FS-7.2 in the simulations. 

The provided parameters are treated as constants in the optimization problems \eqref{eq:optimization1}, \eqref{eq:optimization3}, and  \eqref{eq:sumSE-optimization}. Once the solutions to the optimization problems are found, the set of active O-RUs, the O-RU cluster for each UE, and transmit power allocation are determined by checking the non-zero power coefficients $\rho_{k,l}$ found by each method. The required GOPS can then be computed using the expression \eqref{eq:CGPP} according to the active O-RUs, clusters, and functional splitting option. The number of active LCs and GPPs are set differently for each resource orchestration scheme as explained in detail in Section~\ref{sec:sumrate}.A. In the end, the total power consumption in \eqref{eq:Xobjective} is computed using the fixed parameters, the output of the optimization problems, and the virtualization level of the considered resource orchestration.

In the CCP-based algorithms, five random initializations are considered and the best solution among them is noted. For the non-fully virtualized resource allocation schemes, for each setup, five random fixed TWDM-PON wavelength and LC connections are assumed and the averages of the power consumption values are presented in the figures. For the solution of \eqref{eq:optimization3}, the $\alpha$ parameter in \eqref{eq:concave-approx} is selected as $7-\mathrm{SE}_{\rm req}$, where $\mathrm{SE}_{\rm req}$ is the required SE value for all the UEs and this selection is empirically based on the several experiments and the observation that larger $\alpha$ is needed for smaller SE requirement. For the problem in \eqref{eq:sumSE-optimization}, $\alpha=3$ is selected. For the algorithms, a minimum and a maximum iteration number of 10 and 50 are set, respectively. The solution accuracy parameter in Algorithm~\ref{alg:ccp} and the threshold in Algorithm~\ref{alg:ccp2} are selected as $\varepsilon=10^{-5}$ and $\zeta=10^{-3}$, respectively.

First, we consider a setup with $L=16$ O-RUs, $K=8$ UEs, and $W=4$ GPPs with FS-8 and find the optimal end-to-end resource allocation by solving the mixed-binary second-order cone programming problem in \eqref{eq:optimization1}. In addition to three resource allocation schemes with cell-free massive MIMO, we also consider a conventional small-cell system where each UE is only served by one O-RU with end-to-end resource allocation. To this end, we consider the same virtualized O-RAN architecture and obtain the power-optimal resource allocation by solving \eqref{eq:optimization1} for the small-cell system by adding an additional constraint $\sum_{l=1}^L x_{k,l}=1$, $\forall k$ to guarantee that only one O-RU is transmitting data to each UE.
 Hence, the problem for small-cell is more restrictive than its cell-free counterpart.

\begin{table*}[t]
	\small
	\caption{Simulation Parameters.}  \label{tab:simulation}
		\centering
	\begin{tabular}{|c|c|c|c|c|c|c|}
		\hline
	 $N$&   4    &    
	 $f_s$, $B$ &  30.72\,MHz, 20\,MHz    &
	 
	 $N_{\rm DFT}$, $N_{\rm used}$ & 2048, 1200 \\ \hline
	 $T_s$ & 71.4\,$\mu$s &
	 $N_{\rm smooth}$, $N_{\rm slot}$ & 12, 16 &
	 $\tau_c$, $\tau_p$ & 192, 8 \\ \hline
Size of coverage area &  1\,km $\times$\,1\,km      &
	 $P_{{\rm RU},0}$, \ $\Delta^{\rm tr}$  &  6.8$N$\,W, \ 4  &
	  Pilot power, \ $p_{\rm max}$ & 100\,mW, \   1\,W    \\ \hline
 $P_{\rm fixed}$, \ $\sigma_{\rm cool}$  & 120\,W, \ 0.9 &
  $P_{\rm ONU}$, \ $P_{\rm OLT}$ & 7.7\,W, \ 20\,W  &
 $P_{{\rm GPP},0}^{\rm proc}$, $P_{{\rm RU},0}^{\rm proc}$ & 20.8\,W \\  \hline
 $\Delta^{\rm proc}_{\rm GPP}$, $\Delta^{\rm proc}_{\rm RU}$ & 74\,W &
 $C_{\rm GPP}^{ \rm max}$, $C_{\rm RU}^{ \rm max}$ & 180\,GOPS &
 $R_{\rm max}$, $N_{\rm bits}$ & 10\,Gbps, 12 \\ \hline
\end{tabular}
\end{table*}

\begin{figure}[t!]
		\begin{center}
			\includegraphics[trim={1cm 0.1cm 1.6cm 0.8cm},clip,width=3.3in]{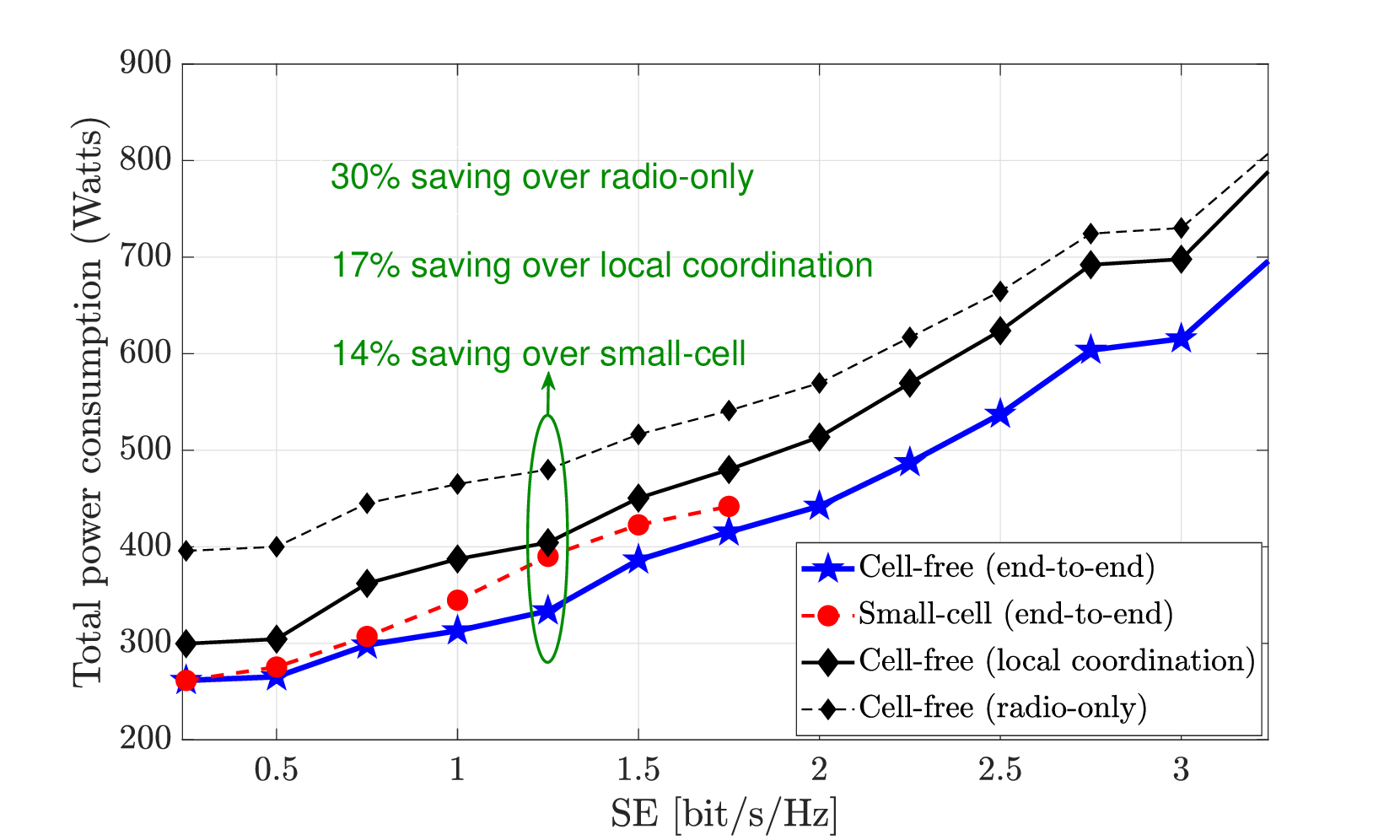}
					\caption{The total power  versus the SE requirement per UE for $L=16$ and $K=8$. } \label{fig:figsim1}
		\end{center}
\end{figure}

We consider 30 random O-RU and UE locations. For each random setup, we take the average of five random permutations of the fixed fronthaul wavelength assignments in the local coordination-based and radio-only resource allocation schemes. Fig.~\ref{fig:figsim1} shows the average total power consumption for a given SE requirement that is assumed to be the same for every UE. We consider the same random setups for all the methods. However, due to the SINR constraints in \eqref{eq:Xconstraint1}, the optimization problem is not guaranteed to be feasible. In the figure, the average is taken out of all feasible setups at each point and it is only plotted when the feasibility ratio is greater than 50\%. As the plot shows, when the SE requirement is greater than 1.75, the small-cell system cannot guarantee reliable performance due to infeasibility. On the other hand, the cell-free system benefits from user-centric JT to support the UEs with much higher SEs. For small SE values, the power consumption is almost the same for both systems. The reason is that for some setups, the cell-free optimization problem results in small-cell solution making their power consumption the same. However, as the SE increases, more O-RUs and DUs are activated to serve UEs when using the small-cell system. This results in increased power consumption compared to cell-free massive MIMO, as shown in the figure for the SE range $[0.5,1.75]$. The maximum power saving is $14\%$ when the SE requirement is 1.25 bit/s/Hz. In conclusion, cell-free massive MIMO results in less or equal power consumption for a small-cell system to guarantee a certain SE requirement. Based on our simulations, we observe that cell-free massive MIMO provides around 1.7 times more rate to the UEs with almost the same energy per bit in comparison to the small-cell system. 

\begin{figure}[t!]
\begin{center}
			\includegraphics[trim={1cm 0.1cm 1.6cm 0.8cm},clip,width=3.3in]{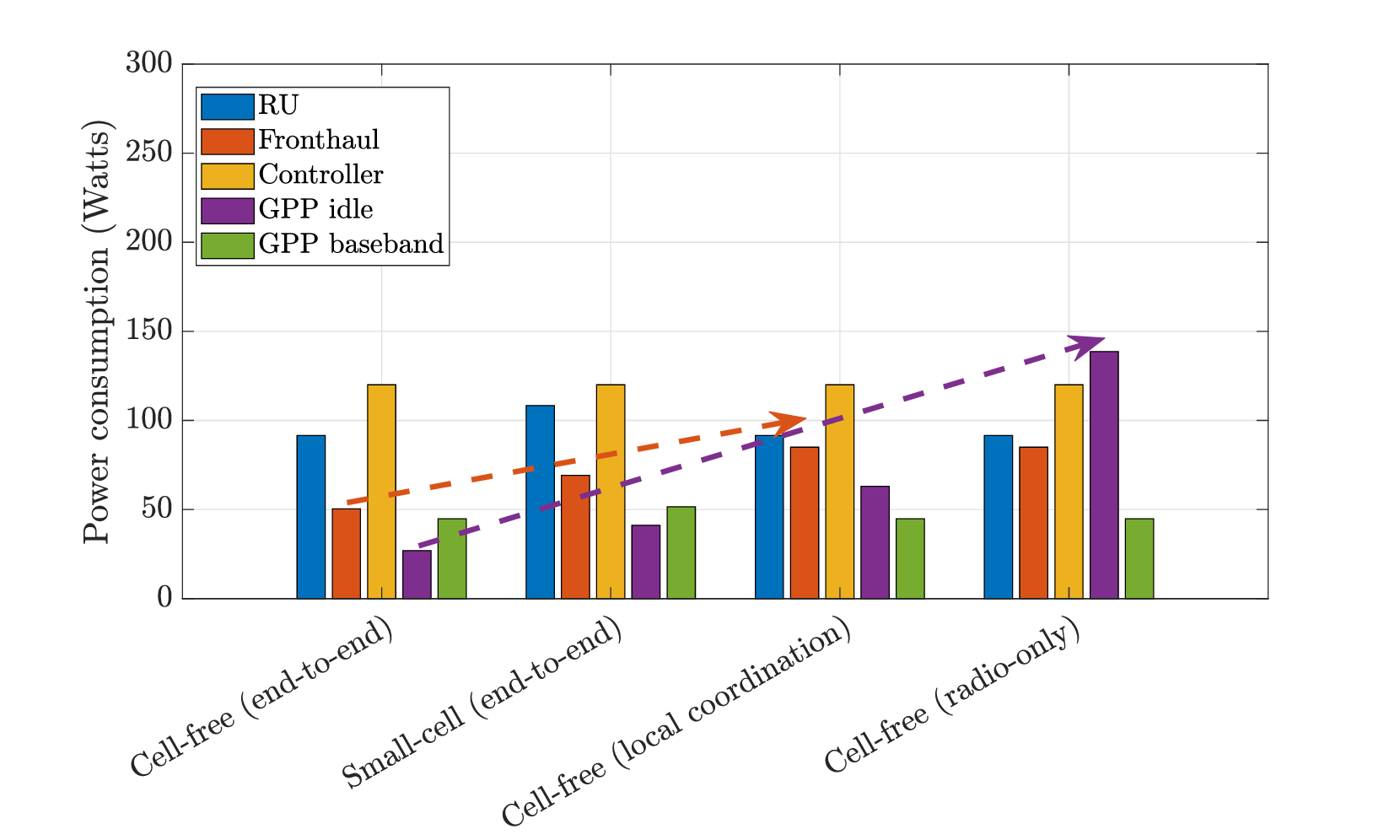}
					\caption{Power consumption breakdown for the SE requirement of $1.25$\,bit/s/Hz, $L=16$, and $K=8$. } \label{fig:figsim2}
		\end{center}
\end{figure}

When we compare the total power consumption of the virtualized cell-free massive MIMO system with the other two orchestration schemes, we see a consistent power saving achieved by the former one for all the considered SE values. 30\% and 17\% saving are obtained when the SE requirement is 1.25\,bit/s/Hz over radio-only and local coordination-based resource allocation, respectively. To better understand what contributes most to the achieved power saving, we show the average power consumption breakdown for the particular SE requirement of 1.25 bit/s/Hz in Fig.~\ref{fig:figsim2}. Virtualized cell-free massive MIMO with end-to-end resource allocation provides reduced power consumption in comparison to the small-cell system due to the reduced RAN, fronthaul, and cloud processing power consumption when cell-free operation is used thanks to the less number of activated network components. When cell-free operation is used with local coordination, the frounthaul and GPP idle power consumption increases due to the fixedly assigned fronthaul resources and the partial intra-LC cloud resource sharing mechanism. The GPP idle power further increases when the radio-only scheme is considered due to the increased number of activated LCs and GPPs.

In Table~\ref{tab:comparison}, we list the total power consumption of fully virtualized cell-free massive MIMO for different SE requirements from Fig.~\ref{fig:figsim1}, which corresponds to the optimal solution found by solving the mixed-binary second-order cone programming problem in \eqref{eq:optimization1}. We compare these optimal values with those obtained by solving the proposed approximate problem in \eqref{eq:optimization3} via the CCP approach outlined in Algorithm~\ref{alg:ccp} and the refinement method in Algorithm~\ref{alg:ccp2}. As the table shows, there is a slight power consumption increase when solving the lower-complexity approximate problem. However, the resulting increase is at most 8\%, which showcases the effectiveness of the proposed method. 

\begin{table*}[h!]
	\small
			\caption{Comparison of the total power consumption (W) between the optimal solution and the CCP approach.}  \label{tab:comparison}
	 \hspace{0.8cm}
	\begin{tabular}{|c|c|c|c|c|c|c|c|c|c|c|c|c|c|}
		\hline
	 SE [bit/s/Hz]&   0.25    &    0.5 & 0.75 & 1 & 1.25 & 1.5 & 1.75 & 2 & 2.25 & 2.5 & 2.75 & 3 & 3.25	\\ \hline
  Optimal solution &  262 & 265 & 298 &
  313 & 334 &   386 &  415 & 442  &
  487 &  537  & 604 & 615 &
699   \\ \hline
  CCP solution &  266 & 282 & 304 &
  333 & 361 &   400 &  432 & 473  &
  517 &  571  & 646 & 666 &
756   \\ \hline
\end{tabular}
\end{table*}

\begin{figure}[t!]
		\begin{center}
			\includegraphics[trim={1cm 0.1cm 1.6cm 0.8cm},clip,width=3.3in]{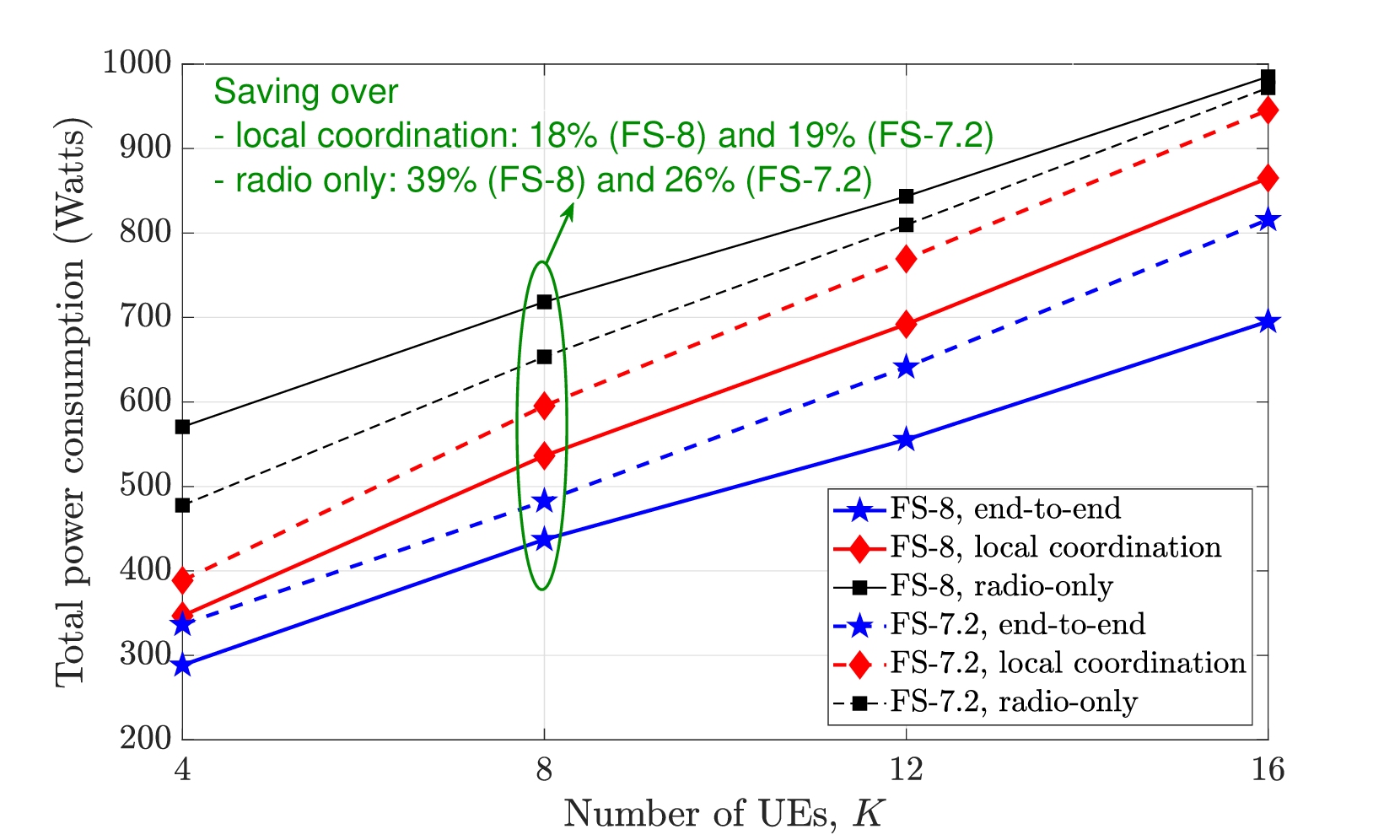}
						\caption{The total power  versus the number of UEs for $L=36$ and the SE requirement of 2\,bit/s/Hz. } \label{fig:figsim3}
		\end{center}
\end{figure}

\begin{figure}[t!]
		\begin{center}
			\includegraphics[trim={1cm 0.1cm 1.6cm 0.8cm},clip,width=3.3in]{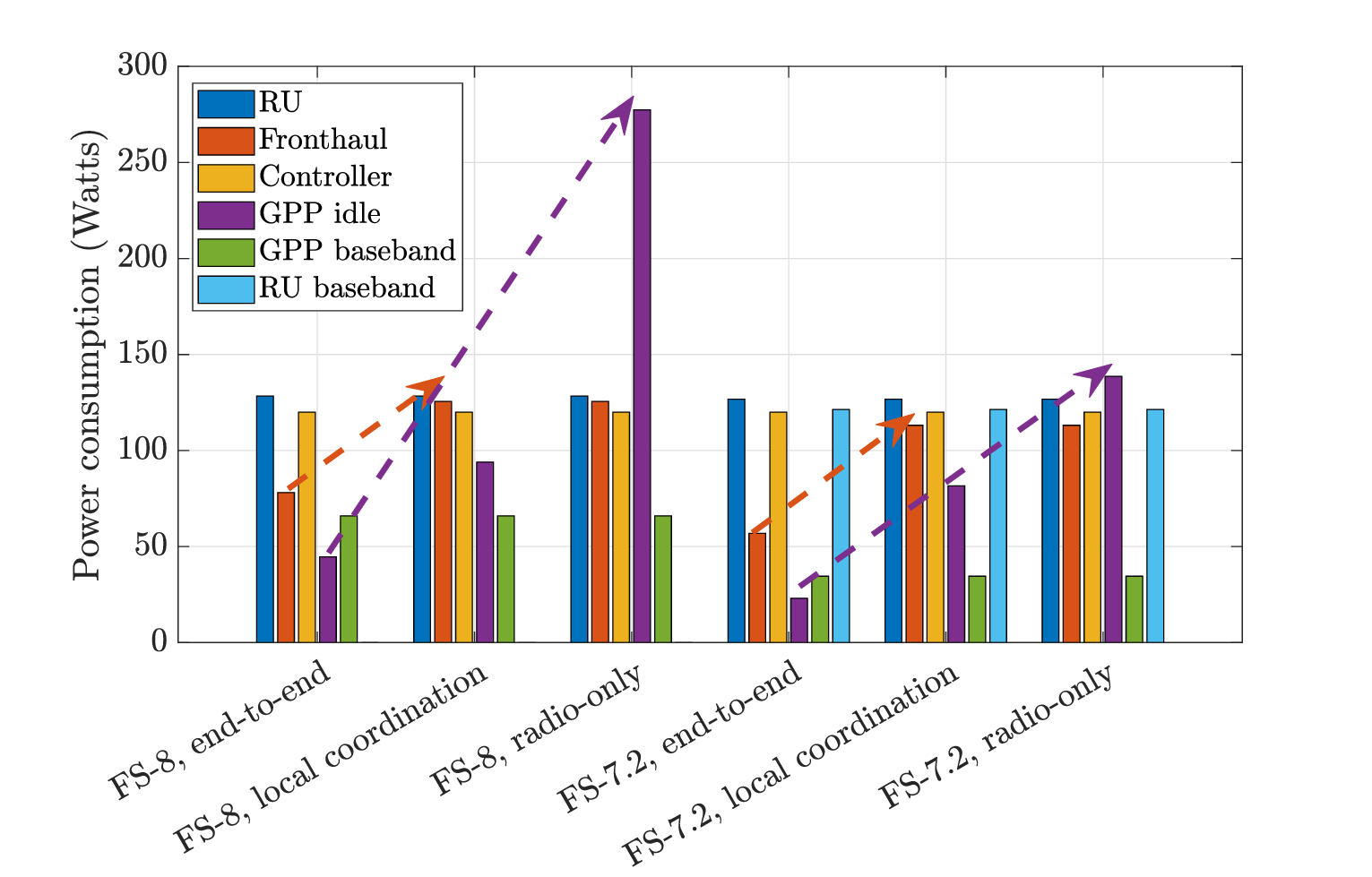}
				\caption{Power consumption breakdown for $K=8$, $L=36$, and the SE requirement of $2$\,bit/s/Hz. } \label{fig:figsim4}
		\end{center}

\end{figure}

In the remainder of the simulations, we consider a setup with $L=36$ O-RUs.  We do not specify the maximum number of GPPs and LCs, just assign the required number of them. First, we solve the proposed problem in \eqref{eq:optimization3} by the CCP approach outlined in Algorithm~\ref{alg:ccp} and the refinement method in Algorithm~\ref{alg:ccp2} for different numbers of UEs, where the  SE requirement for each UE is 2\,bit/s/Hz. As seen in Fig.~\ref{fig:figsim3}, for both FS-8 and FS-7.2, the end-to-end resource allocation results in smaller total power consumption thanks to the pooled cloud resources and sharing among the GPPs. Compared to radio-only orchestration, the fully virtualized end-to-end resource allocation provides 39\% power saving for $K=8$. On the other hand, when the intra-PHY split FS-7.2 is used, a processing unit is activated for each active O-RU leading to inefficient resource utilization with increased power. This reduces the cloud processing requirements under the assumption that all O-RUs are connected to the O-Cloud in the radio-only scheme. Moreover, each LC can now serve two times more O-RUs ($6/3$) than FS-8 allows, thanks to the corresponding significantly reduced fronthaul rate requirements. Hence, the power saving obtained by the fully virtualized orchestration over the radio-only one becomes smaller, i.e., 26\%. This can also be observed by the power consumption breakdown illustrated in Fig.~\ref{fig:figsim4}. When FS-8 is used, there is a substantial saving opportunity regarding the GPP idle power. On the other hand, this opportunity is less apparent when FS-7.2 is used since RU baseband processing power is the same for all three orchestration schemes.

\begin{figure}[t!]

		\begin{center}
			\includegraphics[trim={1cm 0.1cm 1.6cm 0.8cm},clip,width=3.3in]{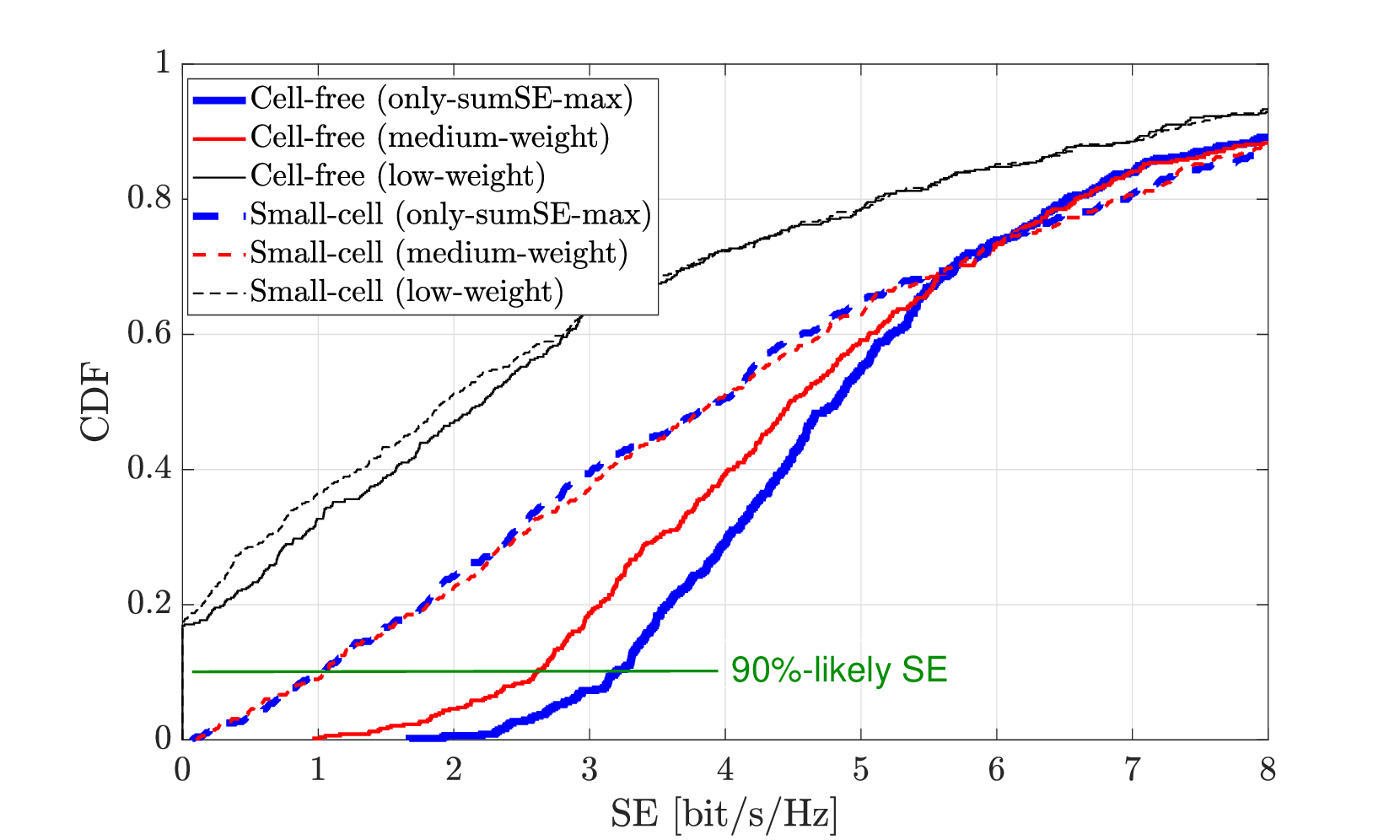}
						\caption{The CDF of SE per UE for the joint sum SE maximization and power consumption minimization problem. For the sake of presentation, the CDF values are only shown for the SE range $[0-8]$\,bit/s/Hz. } \label{fig:figsim5}
		\end{center}
	\end{figure}

\begin{figure}[t!]
		\begin{center}
			\includegraphics[trim={1cm 0.1cm 1.6cm 0.8cm},clip,width=3.3in]{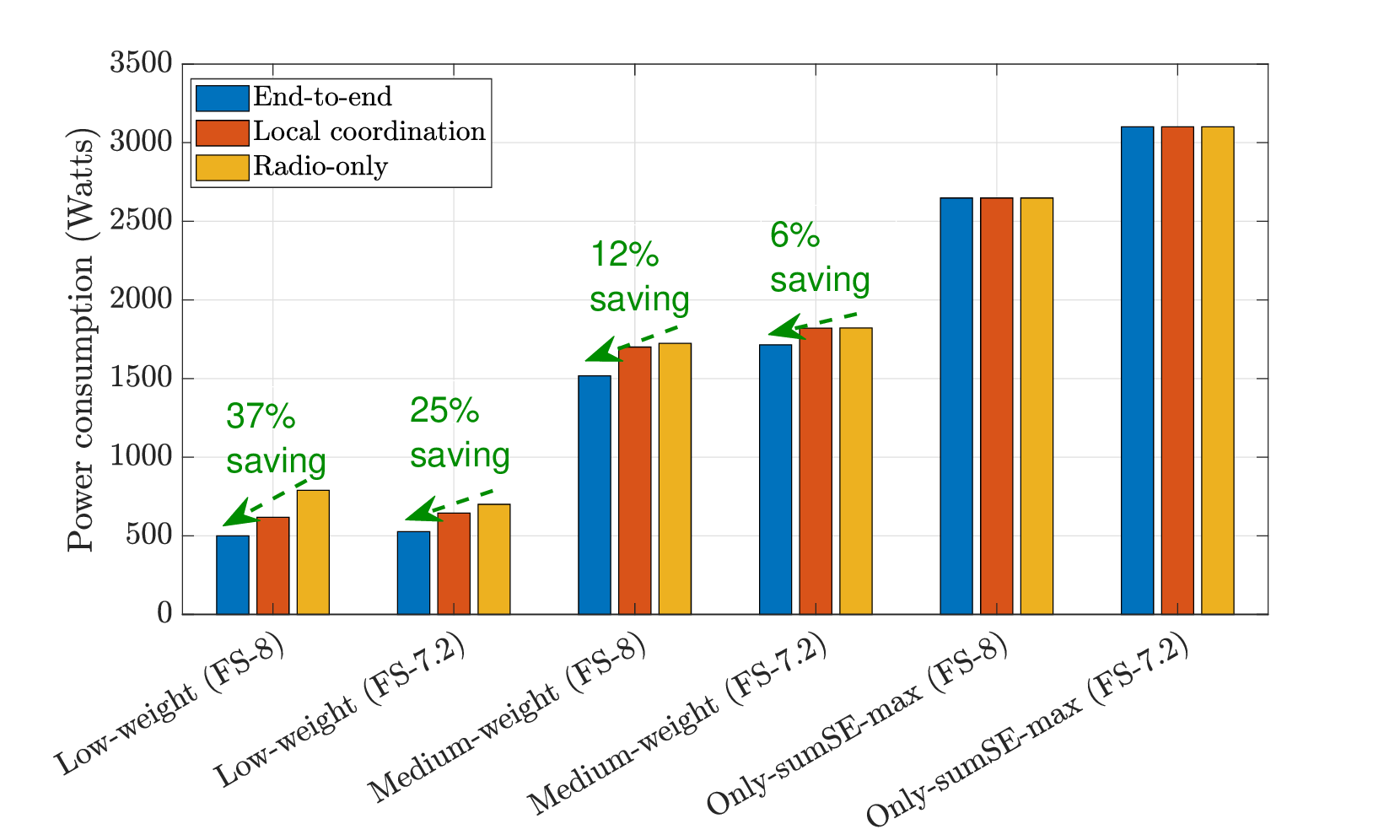}
					\caption{The total power consumption for the joint sum SE maximization and power consumption minimization problem. } \label{fig:figsim6}
		\end{center}
\end{figure}

In Figs.~\ref{fig:figsim5}, \ref{fig:figsim6}, and Table~\ref{tab:simulation2}, we present the results of the joint sum SE maximization and total power consumption minimization problem in \eqref{eq:sumSE-optimization3}, which is solved by the proposed CCP algorithmic framework similar to Algorithm~\ref{alg:ccp}. To obtain the solution of the small-cell system, we select the O-RU that has the largest assigned power by the cell-free solution for each UE. Two values of the penalty parameter $\lambda$ for the sum SE in the multi-objective function in \eqref{eq:sumSE-optimization3} are considered: i) $\lambda=5$, which is called low-weight scenario and ii) $\lambda=50$, which is called medium-weight scenario according to our observations and trials. As another benchmark, we consider only the sum SE maximization approach, where the power consumption is not included in the objective function. In Fig.~\ref{fig:figsim5}, the cumulative distribution function (CDF) of the SE per UE of all the considered three methods for both cell-free and small-cell systems with FS-8. The SE values are very close to the ones in this figure when FS-7.2 is selected; thus, we skip those results. As shown in this figure and Table~\ref{tab:simulation2}, the main benefit of cell-free operation is the significantly increased SE for the most unfortunate UEs with the lowest SEs. The so-called 90\%-likely SE, which can be provided to 90\% of all the UEs, increases by more than two-fold when cell-free massive MIMO transmission is adopted. The sum SE also improves with the cell-free massive MIMO, but its impact is less significant considering the increased end-to-end power consumption. The increased power consumption is mainly compensated for by the higher SE guaranteed to all the UEs  provided by cell-free massive MIMO. As the weight given to the sum SE maximization increases in the problem, this worst-case SE improves but with a cost of higher total power consumption.

In Fig.~\ref{fig:figsim6}, we plot the total power consumption for the three resource allocation schemes and for different weights given to the sum SE and functional splits. It is worth noting that for a given weight, all the resource allocation schemes provide the same SE values. The difference lies in the power saving achieved by full virtualization and resource sharing in the cloud. As shown in the figure, the power saving is higher when the weight is low, and hence, the number of active O-RUs and the respective radio power are in a small value range.  As observed before, the power saving is higher when FS-8 is utilized. As also demonstrated in Table~\ref{tab:simulation2}, the total power consumption is consistently larger with the FS-7.2.


\begin{table*}[t]
	\footnotesize
	\caption{Sum SE, 90\%-likely SE, and total power consumption for the joint sum SE maximization and power consumption minimization problem.}  \label{tab:simulation2}
		\centering
\begin{tabular}{c|ccc|ccc|}
\cline{2-7}
                                 & \multicolumn{3}{c|}{\begin{tabular}[c]{@{}c@{}}Sum SE\\ (bit/s/Hz)\end{tabular}}      & \multicolumn{3}{c|}{\begin{tabular}[c]{@{}c@{}}90\%-likely SE\\ (bit/s/Hz)\end{tabular}} \\ \cline{2-7} 
                                 & \multicolumn{1}{c|}{Low-weight} & \multicolumn{1}{c|}{Medium-weight} & Only-sumSE-max & \multicolumn{1}{c|}{Low-weight}  & \multicolumn{1}{c|}{Medium-weight}  & Only-sumSE-max  \\ \hline
\multicolumn{1}{|c|}{Cell-free}  & \multicolumn{1}{c|}{46.9}          & \multicolumn{1}{c|}{80.5}              &     84.0           & \multicolumn{1}{c|}{0}            & \multicolumn{1}{c|}{2.6}               & 3.2                 \\ \hline
\multicolumn{1}{|c|}{Small-cell} & \multicolumn{1}{c|}{45.3}           & \multicolumn{1}{c|}{71.9}              &  71.3               & \multicolumn{1}{c|}{0}            & \multicolumn{1}{c|}{1.0}               & 1.0                \\ \hline
\end{tabular}
\begin{tabular}{c|cccccc|}
\cline{2-7}
                                 & \multicolumn{6}{c|}{Total power consumption (Watts)}                                                                                                                                               \\ \cline{2-7} 
                                 & \multicolumn{3}{c|}{FS-8}                                                                                  & \multicolumn{3}{c|}{FS-7.2}                                                           \\ \cline{2-7} 
                                 & \multicolumn{1}{c|}{Low-weight} & \multicolumn{1}{c|}{Medium-weight} & \multicolumn{1}{c|}{Only-sumSE-max} & \multicolumn{1}{c|}{Low-weight} & \multicolumn{1}{c|}{Medium-weight} & Only-sumSE-max \\ \hline
\multicolumn{1}{|c|}{Cell-free}  & \multicolumn{1}{c|}{500}          & \multicolumn{1}{c|}{1518}              & \multicolumn{1}{c|}{2649}               & \multicolumn{1}{c|}{527}           & \multicolumn{1}{c|}{1715}              &       3101         \\ \hline
\multicolumn{1}{|c|}{Small-cell} & \multicolumn{1}{c|}{490}           & \multicolumn{1}{c|}{1062}              & \multicolumn{1}{c|}{1189}               & \multicolumn{1}{c|}{515}           & \multicolumn{1}{c|}{1222}              &  1383               \\ \hline
\end{tabular}
\end{table*}

\section{Conclusions}\label{sec:conclusions}
In this paper, we have modeled the end-to-end network power consumption of a cell-free massive MIMO system in the O-RAN architecture with fully centralized and intra-PHY functional splitting options. We have solved the two end-to-end resource allocation problems to find the power-efficient O-RU selection, O-RU/UE association, and the respective virtualized optical fronthaul and cloud resources. The first problem minimizes end-to-end consumption under the SE requests and network resource constraints. The second problem jointly maximizes sum SE and minimizes total power consumption to construct a balanced trade-off between SE performance and power cost. 

The proposed fully virtualized end-to-end resource allocation achieves up to 39\% and 19\% power saving compared to the radio-only and local coordination-based orchestration, which only benefit from the O-RU turn-off mechanism and/or partial resource sharing among the computational tasks of the O-RUs that share the same fixed fronthaul wavelength. The key enabler of this power saving is the reduced fronthaul and GPP idle power consumption thanks to the less number of activated fronthaul wavelengths (LCs) and GPPs in the O-Cloud. 

The considered cell-free system is advantageous over conventional small-cells in the same O-RAN architecture for both maximum rate and minimum power consumption. When the first optimization problem with the SE request constraints is considered, up to $14\%$ power saving is possible. On the other hand, for very small SE requests, the performance of cell-free and small-cell systems are the same, since the cell-free functionality is not activated. Cell-free massive MIMO increases the maximum provided rate by 1.7 with less energy per bit in comparison to small cells. When the required SE is high (more than 1.75 bit/s/Hz), the small-cell scenario may not be feasible, while by activating cell-free massive MIMO, we can obtain a feasible and power-efficient solution. When it comes to sum SE maximization, the main benefit of cell-free operation is the significantly increased 90\%-likely SE (more than two-fold) over a small-cell system. This comes with a  cost of increased power consumption, which is mainly compensated for by the higher guaranteed SE to all UEs rather than a relatively smaller sum SE improvement.

If the power consumption is prioritized over sum SE by setting the penalty parameter to a low value ($\lambda=5$), we can save power up to 69\% in comparison to the medium-weight setup ($\lambda=50$), but the SE performance is  inferior in terms of both sum SE and 90\%-likely SE. Tuning the penalty parameter as in the medium-weight setup, we can save power up to \%45 by only sacrificing the sum rate by \%4.

\bibliographystyle{IEEEtran}
\bibliography{IEEEabrv,refs}

\end{document}